\documentclass[a4paper,12pt]{article}
\usepackage{ascmac}
\usepackage{amsthm}
\usepackage[ruled,vlined]{algorithm2e}
\usepackage{amsmath,amssymb}
\usepackage{type1cm}
\theoremstyle{plain}
\newtheorem{theorem}{Theorem}[section]
\newtheorem{lemma}{Lemma}[section]

\newtheorem{corollary}{Corollary}[section]
\newtheorem{claim}{Claim}[section]
\newtheorem{definition}{Definition}[section]
\newtheorem{property}{Property}[section]
\theoremstyle{definition}
\newtheorem{remark}{Remark}[section]
\newtheorem{example}{Example}[section]
\numberwithin{equation}{section}
\newcommand{\leng}[1]{|#1|}
\newcommand{\msc}[1]{\mbox{{\sc #1}}}
\newcommand{\BQED}{\hfill \hbox{\rule{8pt}{8pt}}}
\newcommand{\pair}[1]{\langle{#1}\rangle}
\newcommand{\QED}{\hfill \hfill$\square$}
\title{{\bf {\Large Capacity-Insensitive Algorithms for Online Facility Assignment 
Problems on a Line}}\thanks{This work was partially supported by joint project 
of Kyoto University and Toyota Motor Corporation, 
titled ``Advanced Mathematical Science for Mobility Society.''}}
\author{{\sc Tsubasa Harada}%
\thanks{Department of Mathematical and Computing Science, 
Tokyo Institute of Technology, 2-12-1 Ookayama, Meguro-ku, 
Tokyo 152-8550, Japan. {\sf harada.t.ak@m.titech.ac.jp}}\and 
{\sc Toshiya Itoh}%
\thanks{Department of Mathematical and Computing Science, 
Tokyo Institute of Technology, 2-12-1 Ookayama, Meguro-ku, 
Tokyo 152-8550, Japan.  {\sf titoh@c.titech.ac.jp}}\and 
{\sc Shuichi Miyazaki}%
\thanks{Graduate School of Information Science, University of Hyogo, 
8-2-1 Gakuennishi-machi, Nishi-ku, Kobe, Hyogo 651-2197, Japan. 
{\sf shuichi@sis.u-hyogo.ac.jp}}}
\date{}
\begin{document}
\maketitle
\noindent {\sf Abstract:}  
In the online facility assignment problem ${\rm OFA}(k,\ell)$, 
there exist $k$ servers with a capacity $\ell \geq 1$ on a metric space 
and a request arrives one-by-one. The task of an online algorithm is to 
irrevocably match a current request with one of the servers with vacancies 
before the next request arrives. 
As special cases for ${\rm OFA}(k,\ell)$, 
we consider ${\rm OFA}(k,\ell)$ {\it on a line\/}, which is denoted by 
${\rm OFAL}(k,\ell)$ and ${\rm OFAL}_{eq}(k,\ell)$, where the latter 
is the case of ${\rm OFAL}(k,\ell)$ with equidistant servers. In this paper, 
we perform the competitive analysis for the above problems. 
As a natural generalization of the greedy algorithm {\sc grdy}, 
we introduce a class of algorithms called 
MPFS (Most Preferred Free Servers) and show 
that any MPFS algorithm has the capacity-insensitive property, 
i.e., for any $\ell \geq 1$, {\sc alg} is $c$-competitive 
for ${\rm OFA}(k,1)$ iff  
{\sc alg} is $c$-competitive for ${\rm OFA}(k,\ell)$. 
By applying the capacity-insensitive 
property of the greedy algorithm {\sc grdy}, we 
derive the matching upper and lower bounds $4k-5$
on the competitive ratio of {\sc grdy} for ${\rm OFAL}_{eq}(k,\ell)$. 
To investigate the capability of MPFS algorithms, 
we show that the competitive ratio of 
any MPFS algorithm {\sc alg} for ${\rm OFAL}_{eq}(k,\ell)$ 
is at least $2k-1$. Then we propose a new 
MPFS algorithm {\sc idas} (Interior Division for Adjacent Servers) 
for ${\rm OFAL}(k,\ell)$ and show that the competitive ratio of 
{\sc idas} for ${\rm OFAL}_{eq}(k,\ell)$ is at most $2k-1$, i.e., 
{\sc idas} for ${\rm OFAL}_{eq}(k,\ell)$ is best possible 
in all the MPFS algorithms.  \medskip
\section{Introduction} \label{sec-introduction}
%
Online optimization (profit maximization or cost minimization) problems are 
real-time computation, in which a sequence of requests is an input,  
each request is given to an online algorithm one-by-one, and an online algorithm 
must decide how to deal with the current request before the next request arrives. 
Once the decision is fixed for the current request, the online algorithm is not 
allowed to change it later. In general, the efficiency of online algorithms is measured 
by competitive analysis which is initiated by Sleator and Tarjan \cite{ST1985}. 
Informally, we say that an online algorithm {\sc alg} is {\it $c$-competitive\/} 
(or the {\it competitive ratio\/} of {\sc alg} is at most $c$) if the cost of output 
by {\sc alg} is at most $c$ times worse than the optimal cost (the formal definition 
will be given in  Section \ref{subsec-notation-terminology}). 

The {\it online metric matching problem\/} is initiated independently by 
Kalyanasundaram and Pruhs \cite{KalP1993} and 
Khuller et al.~\cite{KMV1994} as an online variant of the minimum 
cost bipartite matching problem, and is formulated as follows: 
$k$ servers are located on a given metric space and $k$ requests (on the metric 
space) are given one-by-one in an online manner. The task of an online algorithm 
is to match each request immediately with one of the $k$ servers. 
The cost of matching a request with a server is determined by the distance 
between them. The goal of the problem is to minimize the sum of the costs 
of matching $k$ requests to $k$ distinct servers. 
For this problem, 
Kalyanasundaram and Pruhs \cite{KalP1993} and 
Khuller et al.~\cite{KMV1994} presented a deterministic online algorithm, which is 
called {\it Permutation} \cite{KalP1993}, 
and showed that it is $(2k-1)$-competitive and best possible. 

Later,  Kalyanasundaram and Pruhs \cite{KalP1998} restricted the underlying metric 
space to be a line and introduced a problem referred to as the 
online matching problem on a line. For this restricted problem, 
Kalyanasundaram and Pruhs \cite{KalP1998} conjectured that (i) there exists a 
9-competitive algorithm and 
(ii) the {\it Work Function\/} algorithm \cite{KoutP1995} has 
a constant competitive ratio, but   
(i) and (ii) were disproved in 
\cite{FHK2005} and \cite{KN2004}, respectively. 
There have been 
extensive studies on  
this problem \cite{AFT2018,ABNPS2014,GL2012,NR2017,R2016, R2018}  
and the best upper bound on the competitive ratio \cite{NR2017, R2018} 
is $O(\log k)$, which is achieved by the 
{\sc robust-matching} algorithm \cite{R2016}. 
While the best lower bound on the competitive ratio \cite{FHK2005} 
has been 9.001 for a long time, Peserico and Scquizzato \cite{PS2021} 
drastically improved it to $\Omega(\sqrt{\log k})$. 

As a variant of the online metric matching problem, 
Ahmed et al.~\cite{ARK2020} formulated the {\it online facility assignment\/} 
problem as follows: There exist $k$ servers located 
equidistantly on a line and 
each request appears (one-by-one) 
on the line. Each server has a {\it capacity\/}, which corresponds to 
the possible number of requests that can be matched to the server. 
Ahmed et al.~\cite{ARK2020} showed (with rough proofs) that the greedy  
algorithm {\sc grdy} \cite{KalP1995} is $4k$-competitive 
and the {\it Optimal-fill\/} algorithm is $k$-competitive for any $k > 2$. 
Itoh~et~al.~\cite{IMS2021} also analyzed the competitive 
ratio for small $k\geq 2$, and showed that (i) for $k=2$, the competitive ratio 
of any algorithm is at least 3 and {\sc grdy} is 3-competitive, i.e.,  
{\sc grdy} is best possible for $k=2$, and 
(ii) for $k=3$, $4 $, and $5$, the competitive ratio of any algorithm is at least 
$1+\sqrt{6}>3.449$, $\frac{4+\sqrt{73}}{3}>4.181$, and 
$\frac{13}{3}>4.333$, respectively. Further results on this problem were 
extensively obtained by Satake \cite{S2022}. 
%
\subsection{Our Contributions} \label{subsec-contribution}
%
In this paper, we deal with the online facility assignment problem, where  
each server has a capacity\footnote{~Note that if a capacity of each server 
is 1, then it is equivalent to the online metric matching problem.}, and consider 
the following cases: (case 1) a capacity of each server is 1; 
(case 2) a capacity of each server is not necessarily 1. In general, 
the competitive analysis for (case 2) may be harder than that for 
(case 1).  Optimistically, we expect for an algorithm to have the 
{\it capacity-insensitive\/} property, i.e., 
if it is $c$-competitive for (case 1), then 
it is also $c$-competitive for (case 2). This property 
makes the algorithm design much easier. 

In Section \ref{sec-mpfs}, we introduce the class of MPFS (Most 
Preferred Free Servers) algorithms and show that any MPFS algorithm 
has the capacity-insensitive property (Corollary \ref{cor-mpfs-separable}). 
In Section \ref{sec-faithful}, we formulate the {\it faithful\/} property 
crucial for the competitive analysis in the subsequent discussions. 
In Section \ref{sec-cr-grdy}, we analyze the competitive ratio of {\sc grdy} 
for ${\rm OFAL}_{eq}(k,\ell)$ and  
derive a lower bound $4k-5$ (in Theorem \ref{thm-lower-grdy}) 
and an upper bound $4k-5$ (in Corollary \ref{cor-upper-grdy-eq}). 
In Section \ref{sec-lower-mpfs}, 
we show that for any MPFS algorithm {\sc alg} for ${\rm OFAL}_{eq}(k,\ell)$, 
the competitive ratio of {\sc alg} is at least $2k-1$, while in Section \ref{sec-opt-mpfs}, 
we propose a new MPFS algorithm {\sc idas} (Interior Division for Adjacent Servers) 
for ${\rm OFAL}(k,\ell)$ and show that the competitive ratio of 
{\sc idas} for ${\rm OFAL}_{eq}(k,\ell)$ is at most $2k-1$, i.e., 
{\sc idas} for ${\rm OFAL}_{eq}(k,\ell)$ is best possible 
in all the MPFS algorithms. 
%
\subsection{Related Work} \label{subsec-related}
%
Another version of the online metric matching problem was initiated by 
Karp et al.~\cite{KVV1990}. Since it has application to ad auction, 
several variants of the problem have been  extensively studied (see, e.g., 
\cite{M2012} for a survey). 

For the online metric matching problem with $k$ servers, 
a deterministic algorithm (called Permutation 
algorithm \cite{KalP1993}) is known, which  
is $(2k-1)$-competitive and best possible, but 
probabilistic algorithms with better competitive ratio 
\cite{BBGN2007,MNP2006} 	are also shown. 

Ahmed et al.~\cite{ARK2020} also considered the online facility 
assignment problem on a unweighted graph $G$, and showed 
the competitive ratios of {\sc grdy} and 
Optimal-fill algorithms are $2\leng{E(G)}$ and $\frac{\leng{E(G)}k}{r}$, respectively, 
where $r$ is the radius of $G$. Muttakee et al.~\cite{MAR2020} 
derived the competitive ratios of {\sc grdy} and Optimal-fill algorithms 
for grid graphs and the competitive ratio of the Optimal-fill  algorithm 
for arbitrary graphs. There have been extensive studies for the 
online metric matching problem with {\it delays\/} \cite{EKW2016}, 
in which 
an online algorithm is allowed to deter a decision for the current request 
at the cost of waiting time as a ``time cost.'' The goal of the problem is 
to minimize the sum of total matching cost and total time cost. 
There exist studies for deterministic algorithms 
\cite{AF2018, BKLS2018,BKS2017,ESW2017} and the 
best upper bound on the competitive ratio \cite{AF2018}
is $O(m^{\lg (3/2+\varepsilon)}) 
\approx O(m^{0.59})$, where $m$ is the number of requests. 
There also exist studies for randomized algorithms 
\cite{AACCGKMWW2017,ACK2017,EKW2016,LPWW2018}. 
The best upper bound on the competitive ratio is 
$O(\log n)$ by Azar et al.~\cite{ACK2017} and 
the best lower bound for the competitive ratio 
is $\Omega(\frac{\log n}{\log\log n})$ 
by Ashlagi et al.~\cite{AACCGKMWW2017}. 
%
\section{Preliminaries} \label{sec-preliminary}
%
\subsection{Online Facility Assignment Problem} \label{subsec-ofa}
%
Let $(X,d)$ be a metric space, where $X$ is a (possibly infinite) set of points and 
$d: X \times X \to \mathbb{R}$ is a distance function. 
We use  $S=\{s_{1},\ldots,s_{k}\}$ to denote the set of $k$ servers and use 
$\sigma=r_{1}\cdots r_{n}$ to denote 
a request sequence. For each $1 \leq j \leq k$,  a server 
$s_{j}$ is characterized by the position $p(s_{j}) \in X$ and 
$s_{j}$ has capacity 
$c_{j} \in \mathbb{N}$, i.e., $s_{j}$ can be matched with at most $c_{j}$ requests. 
We assume that $n \leq c_{1}+\cdots+c_{k}$. 
For each $1 \leq i \leq n$, 
a request $r_{i}$ is also characterized by the position $p(r_{i}) \in X$. 

The set $S$ is given to an online algorithm in advance, while requests are given 
one-by-one from $r_{1}$ to $r_{n}$. At any time of the execution of an algorithm, 
a server is called {\it free\/} if the number of requests matched with it is less 
than its capacity, and {\it full\/} otherwise. When a request $r_{i}$ is revealed, 
an online algorithm must match $r_{i}$ with one of free servers. If 
$r_{i}$ is matched with $s_{j}$, the pair $(r_{i},s_{j})$
is added to the current matching and the cost 
$\msc{cost}(r_{i},s_{j})=d(p(r_{i}),p(s_{j}))$ is incurred for this pair.
The cost of the matching is the sum of the costs of all the pairs contained in it.
The goal of online algorithms is to minimize the cost of the final matching. 
We refer to such a problem as the {\it online facility assignment\/} problem 
with $k$ servers and denote it by ${\rm OFA}(k,\{c_{j}\}_{j=1}^{k})$. 
For the case that $c_{1}=\cdots=c_{k}=\ell\geq 1$, it is immediate 
that $n \leq k\ell$ and 
we simply use ${\rm OFA}(k,\ell)$ to denote the online facility assignment 
problem with $k$ servers (of uniform capacity $\ell$). 
%
\subsection{Online Facility Assignment Problem on a Line} \label{subsec-ofal}
%
By setting $X=\mathbb{R}$, we can regard 
the online facility assignment problem with $k$ servers 
as the online facility assignment problem {\it on a line\/} with $k$ servers, and 
we denote such a problem by 
${\rm OFAL}(k,\{c_{j}\}_{j=1}^{k})$ for general capacities and 
${\rm OFAL}(k,\ell)$ for uniform capacities. 
In this case, it is immediate that 
$p(s_{j}) \in \mathbb{R}$ for each $1 \leq j \leq k$ 
and $p(r_{i}) \in \mathbb{R}$ for each $1 \leq i \leq n$. 
Without loss of generality, we assume that $p(s_{1}) < \cdots < p(s_{k})$ and 
let 
\begin{equation}
d_{j}=p(s_{j+1})-p(s_{j}) \label{eq-interval}
\end{equation}
for each $1 \leq j \leq k-1$. 
For the case 
that $d_{1}=\cdots = d_{k-1}=d$ with some constant $d>0$, we use 
${\rm OFAL}_{eq}(k,\{c_{j}\}_{j=1}^{k})$ and ${\rm OFAL}_{eq} (k,\ell)$  
to denote ${\rm OFAL}(k,\{c_{j}\}_{j=1}^{k})$ and 
${\rm OFAL}(k,\ell)$ with equidistant $k$ servers, respectively. 
For the subsequent discussion, we assume without loss of generality 
that $d=1$ for both 
${\rm OFAL}_{eq}(k,\{c_{j}\}_{j=1}^{k})$ and ${\rm OFAL}_{eq} (k,\ell)$. 

In the rest of the paper, we will abuse the notations 
$r_{i} \in \mathbb{R}$ and $s_{j} \in \mathbb{R}$ for ${\rm OFAL}(k,\ell)$ 
instead of $p(r_{i}) \in \mathbb{R}$ and $p(s_{j}) \in \mathbb{R}$, respectively, 
when those are clear from the context. 
%
\subsection{Notations and Terminologies} 
\label{subsec-notation-terminology}
%
For a request sequence $\sigma$, let $\leng{\sigma}$ be 
the length of $\sigma$, 
i.e., $\leng{\sigma}=n$ for $\sigma=r_{1}\cdots r_{n}$. 
For a request sequence $\sigma=r_{1}\cdots r_{n}$ and  
a request sequence $\tau=\tilde{r}_{1}\cdots \tilde{r}_{m}$,  
we use $\sigma \circ \tau$ to denote the concatenation of 
$\sigma$ and $\tau$, i.e., $\sigma\circ \tau=r_{1}\cdots r_{n} 
\tilde{r}_{1}\cdots \tilde{r}_{m}$. 

For ${\rm OFA}(k,\{c_{j}\}_{j=1}^{k})$, let $S=\{s_{1},\ldots,s_{k}\}$ be the 
set of $k$ servers. 
For an (online/offline) algorithm {\sc da} for ${\rm OFA}(k,\{c_{j}\}_{j=1}^{k})$ 
and a request sequence 
$\sigma=r_{1}\cdots r_{n}$,  
we use $s_{\rm da}(r_{i};\sigma|S)$ to denote 
the server with which {\sc da} matches $r_{i}$ for each $1 \leq i \leq n$ when 
{\sc da} processes $\sigma$. 
Let $\msc{da}(r_{i};\sigma|S)$ 
be the cost incurred by {\sc da} to match $r_{i}$ with 
$s_{\rm da}(r_{i};\sigma|S)$, i.e., 
$\msc{da}(r_{i};\sigma|S)=\msc{cost}(r_{i},s_{\rm da}(r_{i};\sigma|S))$. 
For a subsequence $\tau=r_{i_{1}} \cdots r_{i_{m}}$ of 
$\sigma$, we use $\msc{da}(\tau;\sigma|S)$ to denote 
the total cost incurred by {\sc da} to match each  $r_{i_{h}}$ 
with the server $s_{\rm da}(r_{i_{h}};\sigma|S)$, i.e., 
\[
\msc{da}(\tau;\sigma|S)=\sum_{h=1}^{m} \msc{da}(r_{i_{h}};\sigma|S).
\]
When $\tau=\sigma$, we simply write $\msc{da}(\sigma|S)$ instead of 
$\msc{da}(\sigma;\sigma|S)$. 
On defining $s_{\rm da}(r_{i};\sigma|S)$, $\msc{da}(r_{i};\sigma|S)$, 
$\msc{da}(\tau;\sigma|S)$, and 
$\msc{da}(\sigma|S)$, 
it is crucial to indicate the set $S$ of servers explicitly, whose role will become
clear in Theorem \ref{thm-upper-grdy-2} (especially 
in Claims \ref{claim-grdy} and \ref{claim-opt}). 
We~use {\sc opt} to denote the optimal {\it offline\/} algorithm, i.e., 
{\sc opt} knows the entire sequence $\sigma=r_{1}\cdots r_{n}$ 
in advance and {\it minimizes\/} the total cost 
to match each request $r_{i}$ with the server $s_{\rm opt}(r_{i};\sigma|S)$. 
Let {\sc alg} be an online algorithm for ${\rm OFA}(k,\{c_{j}\}_{j=1}^{k})$ and 
$\sigma=r_{1} \cdots r_{n}$ be a request sequence. 
For each $1 \leq i \leq n$, 
we define the {\it type\/} of a request $r_{i}$ w.r.t.~{\sc alg} by 
\[
{\rm type}_{\rm alg}(r_{i}) = 
\pair{s_{\rm alg}(r_{i};\sigma|S),s_{\rm opt}(r_{i};\sigma|S)}. 
\] 

To evaluate the performance of an online algorithm {\sc alg}, 
we use the (strict) competitive ratio. 
We say that {\sc alg} is $c$-competitive if 
$\msc{alg}(\sigma|S) \leq c \cdot \msc{opt}(\sigma|S)$ for 
any request sequence $\sigma$. 
The competitive ratio ${\cal R}(\msc{alg})$ of {\sc alg} is defined 
to be the infimum of $c\geq 1$ such that {\sc alg} is $c$-competitive, i.e., 
${\cal R}(\msc{alg}) = \inf \{c \geq 1: \mbox{{\sc alg} is $c$-competitive}\}$. 
%
\subsection{Technical Lemmas} \label{subsec-technical}
%
As mentioned in Section \ref{subsec-ofa}, the online facility 
assignment problem 
${\rm OFA}(k,\{c_{j}\}_{j=1}^{k})$ is defined by the set 
$S=\{s_{1},\dots,s_{k}\}$ 
of $k$ servers, where the server $s_{j}$ has the capacity $c_{j}$ 
for each $1 \leq j \leq k$, and for any request sequence 
$\sigma=r_{1}\cdots r_{n}$ 
to ${\rm OFA}(k,\{c_{j}\}_{j=1}^{k})$, the condition that 
$n \leq c_{1}+\cdots + c_{k}$ must be met. 

In this subsection, we show that for the design of online algorithms for 
${\rm OFA}(k,\{c_{j}\}_{j=1}^{k})$, 
it is sufficient to deal with the case that $n=c_{1}+\cdots+c_{k}$ 
(in Lemma \ref{lemma-kell}) and 
it is sufficient to deal with the case that 
$c_{1}=\cdots c_{k}=\ell$  (in Lemma \ref{lemma-ell}). 
\begin{lemma} \label{lemma-kell}
For ${\rm OFA}(k,\{c_{j}\}_{j=1}^{k})$, let 
$L = c_{1}+\cdots+c_{k}$. 
For any $c \geq 1$, $\msc{alg}(\sigma|S) \leq c \cdot \msc{opt}(\sigma|S)$ 
for any request sequence $\sigma$ such that $\leng{\sigma} =L$ iff  
$\msc{alg}(\sigma'|S) \leq c \cdot \msc{opt}(\sigma'|S)$ for 
any request sequence $\sigma'$ such that $\leng{\sigma'} \leq L$. 
\end{lemma}
\noindent {\bf Proof:} If 
$\msc{alg}(\sigma'|S) \leq c \cdot \msc{opt}(\sigma'|S)$ for 
any request sequence $\sigma'$ such that $\leng{\sigma'} \leq L$, then 
it is obvious that  
$\msc{alg}(\sigma|S) \leq c \cdot \msc{opt}(\sigma|S)$ for 
any request sequence $\sigma$ such that $\leng{\sigma} =L$. 

We show that if $\msc{alg}(\sigma|S) \leq c \cdot \msc{opt}(\sigma|S)$ for 
any request sequence $\sigma$ such that $\leng{\sigma} =L$, then 
$\msc{alg}(\sigma'|S) \leq c \cdot \msc{opt}(\sigma'|S)$ for 
any request sequence $\sigma'$ such that $\leng{\sigma'}<  L$. 
For a request sequence $\sigma'$ such that $\leng{\sigma'}<L$, 
define 
a request sequence $\sigma$ as follows: 
Append $L-\leng{\sigma'}$ requests at the end of $\sigma'$ to make 
free servers of {\sc opt} {\it full\/} with zero cost. 
Note that $\leng{\sigma}=L$, and 
we have that $\msc{opt}(\sigma'|S)=\msc{opt}(\sigma|S)$ and 
$\msc{alg}(\sigma'|S) \leq \msc{alg}(\sigma|S)$. Thus it follows that 
for any request sequence $\sigma'$ such that $\leng{\sigma'}<L$, 
\[
\msc{alg}(\sigma'|S) \leq \msc{alg}(\sigma|S) \leq c \cdot 
\msc{opt}(\sigma|S) =c \cdot \msc{opt}(\sigma'|S), 
\]
where the 2nd inequality follows from the assumption that 
$\msc{alg}(\sigma|S)\leq 
c \cdot \msc{opt}(\sigma|S)$ for any request sequence 
$\sigma$ such that $\leng{\sigma}=L$. \BQED
\begin{lemma}\label{lemma-ell}
For any $\ell \geq 1$, any $c_{1},\ldots,c_{k}$ such that 
$1 \leq c_{1},\ldots,c_{k} \leq \ell$, and any $c \geq 1$, 
there exists a $c$-competitive algorithm 
for ${\rm OFA}(k,\ell)$ 
iff there exists a $c$-competitive~algorithm for 
${\rm OFA}(k,\{c_{j}\}_{j=1}^{k})$. 
\end{lemma} 
\noindent {\bf Proof:} 
If there exists a $c$-competitive algorithm $\msc{alg}$ for 
${\rm OFA}(k,\{c_{j}\}_{j=1}^{k})$, then 
by setting $c_{1}=\cdots = c_{k}=\ell$, it is obvious that 
$\msc{alg}$ is $c$-competitive for ${\rm OFA}(k,\ell)$. 

We show that if {\sc alg} is $c$-competitive for ${\rm OFA}(k,\ell)$, then 
there exists a $c$-competitive algorithm $\msc{alg}'$ for 
${\rm OFA}(k,\{c_{j}\}_{j=1}^{k})$. 
For each $1 \leq j \leq k$, let $m_{j} = \ell - c_{j}\geq 0$ 
and $\sigma'_{j}$ be  
a sequence of $m_{j}$ requests on $s_{j}$. Let  
$\sigma'=\sigma'_{1} \circ \cdots \circ \sigma'_{k}$. 
Define an online 
algorithm $\msc{alg}'$ for ${\rm OFA}(k,\{c_{j}\}_{j=1}^{k})$ as follows: 
From Lemma \ref{lemma-kell}, it suffices to consider 
a request sequence 
$\sigma$ such that 
$\leng{\sigma} = c_{1}+\cdots+c_{k}$, and 
$\msc{alg}'$ simulates {\sc alg} on $\rho=\sigma'\circ \sigma$. Note that 
$\leng{\rho}=\leng{\sigma'\circ \sigma}= k\ell$, 
and it is immediate that 
\begin{eqnarray*}
\msc{opt}(\sigma|S) & \geq & 
\msc{opt}(\sigma'\circ \sigma|S)=\msc{opt}(\rho|S);\\
\msc{alg}'(\sigma|S) & = & \msc{alg}(\sigma'\circ \sigma|S)=\msc{alg}(\rho|S).
\end{eqnarray*}
Thus it follows that 
for any request sequence $\sigma$ such that $\leng{\sigma}=
c_{1}+\cdots+c_{k}$, 
\begin{eqnarray*}
\msc{alg}'(\sigma|S) & = & \msc{alg}(\sigma'\circ \sigma|S) = \msc{alg}(\rho|S) \\
& \leq & c \cdot \msc{opt}(\rho|S) = c \cdot 
\msc{opt}(\sigma'\circ \sigma|S) \leq c \cdot \msc{opt}(\sigma|S), 
\end{eqnarray*}
where the 1st inequality follows from the assumption that 
$\msc{alg}(\rho|S)\leq 
c \cdot \msc{opt}(\rho|S)$ for any request sequence 
$\rho$ such that $\leng{\rho}=k\ell$. \BQED\medskip%

Based on Lemmas \ref{lemma-kell} and \ref{lemma-ell}, 
we assume that $c_{1}=\cdots=c_{k}=\ell$ and we consider 
only request sequences $\sigma$ 
such that $\leng{\sigma}=k\ell$ 
in the rest of the paper (except for Section \ref{sec-lower-mpfs}). 
\begin{remark} \label{rem-capacity}
Let {\sc da} be an (online/offline) algorithm for ${\rm OFA}(k,\ell)$ and 
$\sigma=r_{1}\cdots r_{n}$ be a request sequence. 
From Lemmas \ref{lemma-kell} and 
\ref{lemma-ell}, 
we assume that 
$\leng{\sigma}/\leng{S} \in \mathbb{N}$ 
denotes the (uniform) capacity of servers in $S$. \QED
\end{remark}
%
\section{Capacity-Insensitive Algorithms} \label{sec-mpfs}
%
In this section, we introduce a novel notion of ``capacity-insensitive algorithms.'' 
We first define a class of MPFS (Most Preferred Free Servers) algorithms. 
\begin{definition} \label{def-mpfs}
Let {\sc alg} be an online algorithm for ${\rm OFA}(k,\ell)$. We say that 
{\sc alg} is an {\sf MPFS (Most Preferred Free Servers)} algorithm if 
for any request sequence $\sigma=r_{1}\cdots r_{n}$ such that $n=k\ell$, 
it behaves as follows: For each $1 \leq i \leq n$, \vspace*{-0.15cm}
%
\begin{enumerate}
\item the priority of servers for $r_{i}$ is determined by only the position of $r_{i}$; 
\vspace*{-0.25cm}
\item {\sc alg} matches $r_{i}$ with a server with the highest priority 
among free servers.
\end{enumerate}
\end{definition}
\noindent 
Let ${\cal MPFS}$ be the class of MPFS algorithms. 
In the subsequent discussion, we show that for any $\msc{alg} \in {\cal MPFS}$
and any $\ell\geq 1$, 
{\sc alg} is $c$-competitive for ${\rm OFA}(k,1)$ 
iff {\sc alg} is $c$-competitive for ${\rm OFA}(k,\ell)$. 
We begin by introducing several ingredients to analyze the properties of 
algorithms in ${\cal MPFS}$. 
\begin{definition} \label{def-partition}
For a request sequence $\sigma$, we say that 
a set $\{\sigma_{i}\}_{i=1}^{m}$ of request 
sequences is a {\sf partition} of $\sigma$ if it satisfies 
the following conditions: \vspace*{-0.15cm}
\begin{enumerate}
\item[(1)] For each $1 \leq i \leq m$, $\sigma_{i}$ is a 
subsequence of $\sigma$; \vspace*{-0.25cm}
\item[(2)] For each $1 \leq i < j \leq m$, 
$\sigma_{i}$ and $\sigma_{j}$ have no common request in $\sigma$; \vspace*{-0.25cm}
\item[(3)] $\leng{\sigma_{1}}+\cdots +\leng{\sigma_{m}}=\leng{\sigma}$.  
\end{enumerate}
\end{definition}
\begin{example} \label{exam-partition}
Let $\sigma=r_{1}r_{2}r_{3}r_{4}r_{5}r_{6}$. 
For $\sigma_{1}=r_{1}r_{2}r_{3}r_{6}$ 
and $\sigma_{2}=r_{4}r_{5}$, 
$\{\sigma_{1}, \sigma_{2}\}$ satisfies the conditions (1), (2), and (3) of Definition 
\ref{def-partition}. Thus 
$\{\sigma_{1},\sigma_{2}\}$ is a partition of $\sigma$. 

For $\tilde{\sigma}_{1}=r_{5}r_{6}$, $\tilde{\sigma}_{2}=r_{2}r_{4}$, and 
$\tilde{\sigma}_{3}=r_{1}r_{3}$, 
$\{\tilde{\sigma}_{1}, \tilde{\sigma}_{2},\tilde{\sigma}_{3}\}$ satisfies 
the conditions (1), (2), and (3) of Definition 
\ref{def-partition}. Thus 
$\{\tilde{\sigma}_{1}, \tilde{\sigma}_{2},\tilde{\sigma}_{3}\}$  is a 
partition of $\sigma$. 

For $\hat{\sigma}_{1}=r_{2}r_{5}$ and $\hat{\sigma}_{2}=r_{1}r_{2}r_{4}r_{6}$, 
$\{\hat{\sigma}_{1},\hat{\sigma}_{2}\}$ does not satisfy 
the condition (2) of Definition \ref{def-partition}. 
Thus $\{\hat{\sigma}_{1},\hat{\sigma}_{2}\}$ 
is {\it not\/} a partition of $\sigma$.  \QED
\end{example}
\begin{definition} \label{def-coprime}
Let {\sc alg} be an online algorithm for ${\rm OFA}(k,\ell)$ and 
$\{\sigma_{i}\}_{i=1}^{\ell}$ be a partition of a request sequence $\sigma$ such that 
$\leng{\sigma}=k\ell$, where $\sigma_{i}=r_{1}^{i} \cdots r_{k}^{i}$ for each 
$1 \leq i \leq \ell$. We say that the partition $\{\sigma_{i}\}_{i=1}^{\ell}$ of $\sigma$ 
is {\sf coprime} w.r.t.~{\sc alg} if 
\[
s_{\rm alg}(r_{s}^{i};\sigma|S) \neq s_{\rm alg}(r_{t}^{i};\sigma|S) \wedge 
s_{\rm opt}(r_{s}^{i};\sigma|S) \neq s_{\rm opt}(r_{t}^{i};\sigma|S) 
\]
for each $1 \leq i \leq \ell$ and any pair 
$1 \leq s,t \leq k$ such that $s\neq t$. 
\end{definition}
%
For a bipartite graph $G=(X\cup Y;E)$, 
we say that $M \subseteq E$ is a {\it matching\/}  
if no vertex is incident to more than one edge in $M$. 
For $\leng{X}= \leng{Y}$, we say that 
a matching $M$ between $X$ and $Y$ is {\it perfect\/} if every vertex  in $X$ 
is incident to an edge in $M$. 
The following theorem 
plays a crucial role to analyze the properties of algorithms 
in ${\cal MPFS}$. 
\begin{theorem}[\mbox{\cite[Corollary 1.57]{HHM2008}}]\label{thm-regular}
For any bipartite graph $G=(X\cup Y;E)$, if $G$~is~$\ell$-regular, 
then $G$ contains a perfect matching. 
\end{theorem}

From Theorem \ref{thm-regular}, we have the following lemma. 
\begin{lemma} \label{lemma-coprime}
Let {\sc alg} be an online algorithm for ${\rm OFA}(k,\ell)$ and $\sigma$ 
be a request sequence such that $\leng{\sigma}=k\ell$. Then 
there exists a coprime partition $\{\sigma_{i}\}_{i=1}^{\ell}$ of $\sigma$ 
w.r.t.~{\sc alg}. 
\end{lemma}
\noindent {\bf Proof:} 
Fix a request sequence $\sigma$ with $\leng{\sigma}=k\ell$ arbitrarily and define 
a bipartite graph $G=(X\cup Y;E)$ as follows: Let $X=Y=\{s_{1},\ldots,s_{k}\}$, 
and $(x,y) \in E$ iff there exists a request $r$ in 
$\sigma$ such that ${\rm type}_{\rm alg}(r)= \pair{x,y}$. 
Since $G$ is $\ell$-regular by construction, we have that $G$ contains a 
perfect matching by Theorem \ref{thm-regular}. 

We show the lemma by induction on $\ell\geq 1$. 
For $\ell=1$, the lemma obviously holds. For any $\ell \geq 2$, 
we assume that the lemma holds for 
$\ell-1$ and we show that the lemma holds for $\ell$. 
For the bipartite graph $G=(X \cup Y;E)$ such that $\leng{X}=\leng{Y}$, 
let $M \subseteq E$ be a perfect matching of $G$. Note that $M$ 
can be represented by a permutation $\pi$ on $\{s_{1},\ldots,s_{k}\}$, i.e., 
$M=\{(s_{i},s_{\pi(i)})\}_{i=1}^{k}$. From the definition of $G$, it follows that 
for each $1 \leq i \leq k$, 
there exists a request $r_{i}^{\ell}$ such that 
${\rm type}_{\rm alg}(r_{i}^{\ell})=\pair{s_{i},s_{\pi(i)}}$. 
Let $\sigma_{\ell}=r_{1}^{\ell}\cdots r_{k}^{\ell}$ and 
define the request sequence $\sigma'$ by deleting $\sigma_{\ell}$ 
from $\sigma$. Then $\sigma'$ can be regarded as a request sequence for 
${\rm OFA}(k,\ell-1)$ and from the induction hypothesis, it follows that 
there exists a 
coprime partition $\{\sigma'_{i}\}_{j=1}^{\ell-1}$ of $\sigma'$ 
w.r.t.~{\sc alg}. Thus $\{\sigma'_{i}\}_{j=1}^{\ell-1} \cup \{\sigma_{\ell}\}$ is a 
coprime partition of $\sigma$ w.r.t.~{\sc alg}, 
and this completes the proof of the lemma. \BQED\medskip%

Informally, an algorithm {\sc alg} 
for ${\rm OFA}(k,\ell)$ is {\it separable\/} if there exists a 
coprime partition 
$\{\sigma_{i}\}_{i=1}^{\ell}$ of $\sigma$ with $\leng{\sigma}=k\ell$ 
such that 
the way of matching servers for $\sigma_{i}$ by {\sc alg} on $\sigma$ 
is completely the same as 
the way of matching servers for $\sigma_{i}$ by {\sc alg} on $\sigma_{i}$ 
for ${\rm OFA}(k,1)$. 
\begin{definition} \label{def-separable}
Let {\sc alg} be an online algorithm for ${\rm OFA}(k,\ell)$. 
For any request sequence $\sigma$ such that $\leng{\sigma}=k\ell$, 
we say that {\sc alg} is {\sf separable} if 
there exists a coprime partition $\{\sigma_{i}\}_{i=1}^{\ell}$ of $\sigma$ 
w.r.t.~{\sc alg} such that for each $1 \leq i \leq \ell$ and each $1 \leq j \leq k$, 
\begin{eqnarray}
s_{\rm alg}(r_{j}^{i};\sigma|S) & = & s_{\rm alg}(r_{j}^{i};\sigma_{i}|S); 
\label{eq-separable-alg}\\
s_{\rm opt}(r_{j}^{i};\sigma|S) & = &  s_{\rm opt}(r_{j}^{i};\sigma_{i}|S), 
\label{eq-separable-opt}
\end{eqnarray}
where $\sigma_{i}=r_{1}^{i}\cdots r_{k}^{i}$. 
\end{definition}

Note that $\leng{\sigma}/\leng{S}=\ell$ and 
$\leng{\sigma_{i}}/\leng{S}=1$  in 
(\ref{eq-separable-alg}) and (\ref{eq-separable-opt}). 
Then from Remark \ref{rem-capacity}, it is immediate that on 
the left hand side of (\ref{eq-separable-alg}) and (\ref{eq-separable-opt}), 
{\sc alg} and {\sc opt} can be regarded as an {\it online\/} algorithm and an 
{\it offline\/} algorithm for ${\rm OFA}(k,\ell)$, respectively, 
and on the right hand side of (\ref{eq-separable-alg}) and 
(\ref{eq-separable-opt}), 
{\sc alg} and {\sc opt} can be regarded as an {\it online\/} algorithm 
and an {\it offline\/} algorithm for ${\rm OFA}(k,1)$, respectively. 
The following lemma plays a crucial role to discuss the properties 
of algorithms in ${\cal MPFS}$. 
\begin{lemma} \label{lemma-separable}
Let {\sc alg} be a {\sf separable} online algorithm for ${\rm OFA}(k,\ell)$. 
For any $\ell\geq 1$, if {\sc alg} is $c$-competitive for ${\rm OFA}(k,1)$, then 
{\sc alg} is $c$-competitive for ${\rm OFA}(k,\ell)$. 
\end{lemma}
\noindent {\bf Proof:} Fix a request sequence $\sigma$ such that 
$\leng{\sigma}=k\ell$ arbitrarily. Then from the assumption that 
{\sc alg} is separable for ${\rm OFA}(k,\ell)$, there exists a  
coprime partition $\{\sigma_{i}\}_{i=1}^{\ell}$ of $\sigma$ w.r.t.~{\sc alg} 
that satisfies (\ref{eq-separable-alg}) and (\ref{eq-separable-opt}), 
where $\sigma_{i}=r_{1}^{i}\cdots r_{k}^{i}$ for each $1 \leq i \leq \ell$.  Then 
\begin{eqnarray*}
\msc{alg}(\sigma|S) & = & \msc{alg}(\sigma;\sigma|S)\\
&  = & 
\sum_{i=1}^{\ell} \sum_{j=1}^{k} \msc{alg}(r_{j}^{i};\sigma|S)
= \sum_{i=1}^{\ell} \sum_{j=1}^{k} \msc{alg}(r_{j}^{i};\sigma_{i}|S)
= \sum_{i=1}^{\ell} \msc{alg}(\sigma_{i}|S);\\
\msc{opt}(\sigma|S) & = & \msc{opt}(\sigma;\sigma|S)\\
&  = & 
\sum_{i=1}^{\ell} \sum_{j=1}^{k} \msc{opt}(r_{j}^{i};\sigma|S)
= \sum_{i=1}^{\ell} \sum_{j=1}^{k} \msc{opt}(r_{j}^{i};\sigma_{i}|S)
= \sum_{i=1}^{\ell} \msc{opt}(\sigma_{i}|S).
\end{eqnarray*} 
Since {\sc alg} is $c$-competitive for ${\rm OFA}(k,1)$, we have that 
\[
\msc{alg}(\sigma|S)=\sum_{i=1}^{\ell} \msc{alg}(\sigma_{i}|S) \leq 
\sum_{i=1}^{\ell} c \cdot \msc{opt}(\sigma_{i}|S) =
c \cdot \sum_{i=1}^{\ell} \msc{opt}(\sigma_{i}|S) = c\cdot \msc{opt}(\sigma|S), 
\]
and this implies that {\sc alg} is $c$-competitive for ${\rm OFA}(k,\ell)$ 
for any $\ell\geq 1$. \BQED\medskip%

The following theorem is one of the main results that captures the 
crucial property of algorithms in ${\cal MPFS}$ 
and plays an important role in the subsequent discussions. 
\begin{theorem} \label{thm-mpfs-separable}
If {\sc alg} for ${\rm OFA}(k,\ell)$ is in ${\cal MPFS}$, then {\sc alg} is separable. 
\end{theorem}
\noindent {\bf Proof:} Fix an arbitrary 
algorithm $\msc{alg} \in {\cal MPFS}$ 
for ${\rm OFA}(k,\ell)$ and a request sequence $\sigma$ such 
that $\leng{\sigma}=k\ell$. 
From Lemma \ref{lemma-coprime}, it follows that there exists a coprime partition 
$\{\sigma_{i}\}_{i=1}^{\ell}$ of $\sigma$ w.r.t.~{\sc alg}, where 
$\sigma_{i}=r_{1}^{i}\cdots r_{k}^{i}$ is a subsequence of $\sigma$ for 
each $1 \leq i \leq \ell$. To complete the proof of the theorem, 
it suffices to show the following two facts: \vspace*{-0.15cm} 
\begin{enumerate}
\item[(1)] there exists an optimal offline algorithm {\sc opt} such that 
$s_{\rm opt}(r_{j}^{i};\sigma|S)=s_{\rm opt}(r_{j}^{i};\sigma_{i}|S)$
for each $1 \leq i \leq \ell$ and each $1 \leq j \leq k$; \vspace*{-0.25cm} 
\item[(2)] $s_{\rm alg}(r_{j}^{i};\sigma|S)=s_{\rm alg}(r_{j}^{i};\sigma_{i}|S)$ 
for each $1 \leq i \leq \ell$ and each $1 \leq j \leq k$. 
\end{enumerate}

For the fact (1), it is obvious that 
$\msc{opt}(\sigma_{i}|S) = \msc{opt}(\sigma_{i};\sigma_{i}|S) 
\leq \msc{opt}(\sigma_{i};\sigma|S)$ for each $1 \leq i \leq \ell$. 
Assume that there exists an $1 \leq h \leq \ell$ 
such that $\msc{opt}(\sigma_{h}|S) = \msc{opt}(\sigma_{h};\sigma_{h}|S) 
< 
\msc{opt}(\sigma_{h};\sigma|S)$. Define 
a subsequence $\sigma-\sigma_{h}$ of $\sigma$ by deleting $\sigma_{h}$ 
from $\sigma$. Then 
\[
\msc{opt}(\sigma|S) = \msc{opt}(\sigma_{h};\sigma|S)
+\msc{opt}(\sigma-\sigma_{h};\sigma|S) >\msc{opt}(\sigma_{h};\sigma_{h}|S)  
+\msc{opt}(\sigma-\sigma_{h};\sigma|S), 
\]
and this contradicts the optimality of {\sc opt}. Thus we have that 
$\msc{opt}(\sigma_{i}|S)=\msc{opt}(\sigma_{i};\sigma|S)$  
for each $1 \leq i \leq \ell$, 
which is achieved 
in such a way that {\sc opt} for ${\rm OFA}(k,\ell)$ matches 
$r_{j}^{i}$ with $s_{\rm opt}(r_{j}^{i};\sigma_{i}|S)$, i.e., 
$s_{\rm opt}(r_{j}^{i};\sigma|S)=s_{\rm opt}(r_{j}^{i};\sigma_{i}|S)$
for each $1 \leq i \leq \ell$ and each $1 \leq j \leq k$. 

We turn to show the fact (2). For simplicity, let 
$s_{\rm alg}(r_{j}^{i};\sigma|S)=s_{j}^{i}$ for each $1 \leq i \leq \ell$ and 
$1 \leq j \leq k$. From the definition of coprime partition, it follows that 
$\{s_{1}^{i},\ldots,s_{k}^{i}\}=S$ for each $1 \leq i \leq \ell$. 
After {\sc alg} matches $r_{j}^{i}$ with $s_{j}^{i}$, 
{\sc alg} matches $r_{j+1}^{i},\ldots,r_{k}^{i}$ with 
$s_{j+1}^{i},\ldots,s_{k}^{i}$, respectively, 
and this implies that $s_{j}^{i},\ldots,s_{k}^{i}$ are {\it free\/} 
just before {\sc alg} matches $r_{j}^{i}$ with $s_{j}^{i}$. 
Since $\msc{alg} \in {\cal MPFS}$, we have that 
$s_{j}^{i}$ has the highest priority for $r_{j}^{i}$ among 
$s_{j}^{i},\ldots,s_{k}^{i}$. As mentioned in Remark \ref{rem-capacity}, 
we regard {\sc alg} as an algorithm for ${\rm OFA}(k,1)$ 
for the request sequence $\sigma_{i}$. When processing $r_{j}^{i}$, 
it is immediate that $s_{1}^{i},\ldots,s_{j-1}^{i}$ are {\it full\/} and 
$s_{j}^{i},\ldots,s_{k}^{i}$ are {\it free\/}. Thus from the fact that 
$s_{j}^{i}$ has the highest priority for $r_{j}^{i}$ among 
free servers 
$s_{j}^{i},\ldots,s_{k}^{i}$, it follows that {\sc alg} for ${\rm OFA}(k,1)$ 
matches $r_{j}^{i}$ with $s_{j}^{i}$. \BQED\medskip%

Then 
we have the following immediate 
corollary to Theorem \ref{thm-mpfs-separable}. 
\begin{corollary} \label{cor-mpfs-separable} 
Let $\msc{alg} \in {\cal MPFS}$. 
For any $c \geq 1$ and any $\ell \geq 1$, 
{\sc alg} is $c$-competitive for ${\rm OFA}(k,1)$ iff 
{\sc alg} is $c$-competitive for ${\rm OFA}(k,\ell)$.
\end{corollary}
\noindent {\bf Proof:} It is obvious that  
{\sc alg} for ${\rm OFA}(k,\ell)$ is $c$-competitive for any $\ell \geq 1$, 
then {\sc alg} for ${\rm OFA}(k,1)$ is $c$-competitive. The converse follows from 
Lemma \ref{lemma-separable} and Theorem \ref{thm-mpfs-separable}.  \BQED
%
\section{Faithful Algorithms} \label{sec-faithful}
%
In this section, we introduce a notion of {\it faithful\/} algorithms, 
which will play a crucial role to analyze upper bounds on the 
competitive ratio of algorithms in ${\cal MPFS}$. Before discussing 
the faithful algorithms, we introduce {\it tours\/} for a set of fixed points 
on a line and we also observe the related notions and properties. 
%
\subsection{Tours and Their Properties} \label{subsec-tour}
%
Let $V=\{v_{1},\ldots,v_{n}\}$ be a set of 
distinct $n$ points (on a line), i.e., $v_{1},\ldots,v_{n} \in \mathbb{R}$. 
We say that 
$T:v_{1}\to v_{2}\to \cdots \to v_{n} \to v_{1}$ is a {\it tour\/} on $V$ and 
define the {\it length\/} of $T$ by 
\[ \ell(T) = \leng{v_{n}-v_{1}}+\sum_{i=1}^{n-1} \leng{v_{i}-v_{i+1}}. 
\]
For each $2 \leq i \leq n$, 
we identify $v_{1}\to \cdots \to v_{n} \to v_{1}$ with  
$v_{i}\to \cdots \to v_{n} \to v_{1} \to \cdots \to v_{i}$.  
\begin{definition} \label{def-conflicting-tour}
Let $T:v_{1}\to \cdots \to v_{n}\to v_{1}$ be a tour on 
$V=\{v_{1},\ldots,v_{n}\}$. We say that a pair $(v_{i},v_{j})$ is {\sf conflicting} in $T$ 
if $v_{i} \leq v_{j+1}<v_{i+1}\leq v_{j}$. 
\end{definition}
\begin{definition} \label{def-relay/turning}
Let $T:v_{1}\to \cdots \to v_{n}\to v_{1}$ be a tour on 
$V=\{v_{1},\ldots,v_{n}\}$. 
We say that $v_{i}$ is a {\sf relay point} in $T$ if 
$v_{i-1} < v_{i} < v_{i+1}$ or $v_{i-1} > v_{i} > v_{i+1}$, 
where $v_{0}=v_{n}$ for $i=1$ 
and $v_{n+1}=v_{1}$ for $i=n$, 
and say that $v_{i}$ 
is a {\sf turning point} in $T$ otherwise.
\end{definition}
For a tour $T$ on $V$, we use ${\rm cf}(T)$ to denote the set of all 
conflicting pairs in $T$ and use 
${\rm tp}(T)$ to denote the set of all turning points in $T$. 
\begin{remark} \label{rem-contracted-even}
$\leng{{\rm tp}(T)}$ is even 
for any tour $T$ on $V=\{v_{1},\ldots,v_{n}\}$. \QED
\end{remark}

\begin{definition} \label{def-contracted-tour}
Let $T:v_{1}\to \cdots \to v_{n}\to v_{1}$ be a tour on 
$V=\{v_{1},\ldots,v_{n}\}$. 
We say that $\tilde{T}: t_{1} \to \cdots \to t_{2m}\to t_{1}$ 
is a {\sf contracted tour} of $T$ if $\tilde{T}$ consists of 
all the turning points in $T$ by skipping 
all the relay points in $T$. 
\end{definition}

For the contracted tour $\tilde{T}$ of $T$, it is immediate that 
$\leng{{\rm tp}(T)}=\leng{{\rm tp}(\tilde{T})}$. Note that~conflicting pairs in $\tilde{T}$ can be defined in a way 
similar to the conflicting pairs in $T$. 
Let ${\rm cf}(\tilde{T})$ be the set of all 
conflicting pairs in $\tilde{T}$. 
\begin{remark}	\label{rem-even-odd}
For a tour $T: v_{1} \to \cdots \to v_{n} \to v_{1}$ on $V$, 
let $\tilde{T}: t_{1}\to \cdots \to t_{2m} \to t_{1}$ be a contracted tour of $T$. 
Then for a conflicting pair $(t_{i},t_{j}) \in {\rm cf}(\tilde{T})$, $i$ is 
even iff $j$ odd. \QED
\end{remark}

For a tour $T: v_{1} \to \cdots \to v_{n} \to v_{1}$, 
let $\tilde{T}: t_{1}\to \cdots \to t_{2m} \to t_{1}$~be~the~contracted~tour 
of $T$. For each $1 \leq p \leq 2m$, 
let $T^{p}:t_{p}=v_{1}^{p} \to \cdots \to v_{x}^{p}=t_{p+1}$ be 
the path from~$t_{p}$ to $t_{p+1}$ in $T$, 
where $t_{2m+1}=t_{1}$, and 
${\rm relay}(t_{p})=\{v_{2}^{p},\ldots,v_{x-1}^{p}\}$ 
be the set of relay points~on~$T^{p}$. 
\begin{remark} \label{rem-relay-points}
${\rm relay}(t_{p})\cap {\rm relay}(t_{q})
=\emptyset$ for each $1 \leq p<q\leq 2m$. \QED
\end{remark}
\begin{lemma} \label{lemma-cf-T/contT}
For a tour $T$, let $\tilde{T}$ be the contracted tour of  
$T$. Then there exists~an~injection $f_{\rm inj}: {\rm cf}(\tilde{T}) \to {\rm cf}(T)$. 
\end{lemma}
\noindent {\bf Proof:} 
For a conflicting pair $(t_{i},t_{j}) \in {\rm cf}(\tilde{T})$, 
it is immediate that $t_{i}\leq t_{j+1}<t_{i+1}\leq t_{j}$ by definition. 
Let $T^{i}: t_{i}=v_{1}^{i}\to \cdots \to v_{x}^{i}=t_{i+1}$ 
be the path from $t_{i}$ to $t_{i+1}$ in $T$ and 
$T^{j}: t_{j}=v_{1}^{j}\to \cdots \to v_{y}^{j}=t_{j+1}$  be the 
path from $t_{j}$ to $t_{j+1}$ in $T$. 
Determine the maximum $1 \leq \alpha < x$ 
such that $v_{\alpha}^{i} \leq t_{j+1}$ and 
the maximum $1 \leq \beta < y$ such that 
$v_{\alpha+1}^{i} \leq v_{\beta}^{j}$. Then 
\[
v_{\alpha}^{i} \leq t_{j+1} \leq v_{\beta+1}^{j} < v_{\alpha+1}^{i}  \leq v_{\beta}^{j}, 
\]
and this implies that $(v_{\alpha}^{i},v_{\beta}^{j}) \in {\rm cf}(T)$. 
Let $f_{\rm inj}: (t_{i},t_{j}) \mapsto (v_{\alpha}^{i},v_{\beta}^{j})$. From Remark 
\ref{rem-relay-points}. it follows that $f_{\rm inj}(t_{i},t_{j}) \neq 
f_{\rm inj}(t_{p},t_{q})$ for 
$(t_{i},t_{j})\neq (t_{p},t_{q})$. \BQED
\begin{definition} \label{def-detour}
For the  contracted tour $\tilde{T}:t_{1}\to \cdots \to t_{2m}\to t_{1}$ 
of a tour $T$ with $m \geq 2$, 
we say that a path $D_{i}:t_{i}\to t_{i+1}\to t_{i+2}\to t_{i+3}$ is a 
{\sf detour} in $\tilde{T}$ if 
\[ t_{i}\leq t_{i+2} < t_{i+1} \leq t_{i+3} \mbox{~~or~~} 
t_{i} \geq t_{i+2}  > t_{i+1}  \geq t_{i+3}, 
\]
where $i+j=i+j-2m$ if $i+j > 2m$ for each $1 \leq j \leq 3$. 
\end{definition}

The following guarantees that a contracted tour $\tilde{T}$ 
of any tour $T$ has a detour in $\tilde{T}$. 
\begin{lemma} \label{lemma-detour}
For a tour $T$, 
let $\tilde{T}:t_{1}\to \cdots \to t_{2m}\to t_{1}$ be the contracted tour of 
$T$. If $m\geq 2$, then there exists a detour in $\tilde{T}$. 
\end{lemma} 
\noindent {\bf Proof:} Without loss of generality, 
we assume that 
$t_{1}=\min \{t_{1},\ldots,t_{2m}\}$ by the definition of contracted tours.
Since $t_{1},\ldots,t_{2m}$ are 
turning points in $\tilde{T}$, 
we have that $t_{1} < t_{3} < t_{2}$ and $t_{3}<t_{4}$.
If $t_{2}\leq t_{4}$, then the path $t_{1}\to t_{2}\to t_{3}\to t_{4}$ 
is a detour $D_{1}$ in $\tilde{T}$ and 
the lemma holds. If $t_{2}> t_{4}$,~then 
$t_{1} < t_{3} < t_{4} <t_{2}$. 
By continuing this process, assume that we reach to the setting that 
\[
t_{1} < \cdots < t_{2i-1} < t_{2i} < \cdots < t_{2}. 
\]
From the definition of turning points in $\tilde{T}$, it follows that 
$t_{2i+1}<t_{2i}$ and $t_{2i+1}<t_{2i+2}$. 
If $t_{2i+1}\leq t_{2i-1}$, then 
$t_{2i-2}\to t_{2i-1}\to t_{2i}\to t_{2i+1}$ is a 
detour $D_{2i-2}$ in $\tilde{T}$ and the lemma~holds. 
If $t_{2i+1}> t_{2i-1}$, then it is immediate that 
\[
t_{1} < \cdots < t_{2i-1} < t_{2i+1} < t_{2i} < \cdots < t_{2}. 
\]
If $t_{2i}\leq t_{2i+2}$, then 
$t_{2i-1}\to t_{2i}\to t_{2i+1}\to t_{2i+2}$ is a detour $D_{2i-1}$
in $\tilde{T}$ and the lemma holds. If $t_{2i}> t_{2i+2}$, then 
we reach to the setting that 
\[
t_{1} < \cdots < t_{2i-1} < t_{2i+1} < t_{2i+2} < 
t_{2i} < \cdots < t_{2}. 
\] 
As a result, we reach to the final setting that 
$t_{1} < \cdots < t_{2m-1} < t_{2m} < \cdots < t_{2}$, 
however, 
$t_{2m-2}\to t_{2m-1}\to t_{2m}\to t_{1}$ is a detour $D_{2m-2}$ 
in  $\tilde{T}$ and the lemma holds. Thus there always exists a detour 
in $\tilde{T}$, 
and this completes the proof of the lemma. \BQED\medskip

We have the following result on conflicting pairs and 
turning points in a tour $T$. 
\begin{lemma} \label{lemma-conflicting}
Let $T: v_{1}\to \cdots \to v_{n} \to v_{1}$ be a tour 
on $V=\{v_{1},\ldots,v_{n}\}$. Then 
\[
\leng{{\rm cf}(T)} \geq \frac{\leng{{\rm tp}(T)}}{2}.
\]
\end{lemma}
\noindent {\bf Proof:} Let $\tilde{T}: t_{1}\to \cdots \to t_{2m}\to t_{1}$ 
be the contracted tour of $T$. 
Since $\leng{{\rm cf}(T)} \geq \leng{{\rm cf}(\tilde{T})}$ 
by Lemma \ref{lemma-cf-T/contT}, 
it suffices to show that 
$\leng{{\rm cf}(\tilde{T})}\geq m$.
We show this by induction on $m\geq 1$. 
For the case that $m=1$, it follows  that $(t_{1},t_{2})$ is a 
conflicting pair in $\tilde{T}$. 

Assume that the lemma holds for $m \geq 1$, i.e., 
for any contracted tour $\tilde{T}$ with~$2m$~turning points, 
there exist at least $m$ 
conflicting pairs in $\tilde{T}$. Let 
$\tilde{T}':\tau_{1}\to \cdots \to \tau_{2(m+1)}\to \tau_{1}$ be the  
contracted tour of a tour $T'$. 
From Lemma \ref{lemma-detour}, there exists a detour 
$\tau_{i}\to \tau_{i+1}\to \tau_{i+2}\to \tau_{i+3}$ in $\tilde{T}'$. 
For the contracted tour $\tilde{T}'$, 
define $\tilde{T}^{*}$ by replacing the detour 
$\tau_{i}\to \tau_{i+1}\to \tau_{i+2}\to \tau_{i+3}$ in $\tilde{T}'$ 
with the arrow $\tau_{i}\to \tau_{i+3}$, i.e., 
$\tilde{T}^{*}:\tau_{1}\to \cdots \to \tau_{i} \to \tau_{i+3}\to 
\cdots \to \tau_{2(m+1)} \to \tau_{1}$. 
Note that $\tilde{T}^{*}$ consists of 
$2m$ turning points. 
Then~by~the~induction hypothesis, 
there exist at least $m$ conflicting pairs in $\tilde{T}^{*}$. 
Since $\tilde{T}^{*}$ loses a conflicting pair 
$(\tau_{i},\tau_{i+1})$ in $\tilde{T}'$, we have that 
there exist at least $m+1$ 
conflicting pairs in $\tilde{T}'$. \BQED
%
\subsection{Faithful Algorithms and Opposite Request Sequences} 
\label{subsec-faithful-opposite}
%
In this subsection, we introduce a notion of {\it faithful\/} algorithms and 
a notion of {\it opposite\/} sequences, which will make 
the competitive analysis easier. 

\begin{definition} \label{def-close}
Let {\sc alg} be an online/offline algorithm for ${\rm OFAL}(k,\ell)$ and 
$\sigma=r_{1}\cdots r_{k\ell}$ and $\tau=q_{1}\cdots q_{k\ell}$ be 
request sequences. We say that $\tau$ is 
{\sf closer} to $S$ than $\sigma$ w.r.t.~{\sc alg} if \vspace*{-0.25cm}
\begin{enumerate}
\item for each $1 \leq i \leq k\ell$, $q_{i}$ is not farther than $r_{i}$ to 
$s_{\rm alg}(r_{i};\sigma|S)$ with which {\sc alg} matches $r_{i}$, i.e., 
$r_{i} \geq q_{i} \geq s_{\rm alg}(r_{i};\sigma|S)$ or 
$r_{i} \leq q_{i} \leq s_{\rm alg}(r_{i};\sigma|S)$. \vspace*{-0.25cm}
\item there exists $1 \leq i \leq k\ell$ such that 
$q_{i}$ is closer than $r_{i}$ to  
$s_{\rm alg}(r_{i};\sigma|S)$ with which {\sc alg} matches $r_{i}$, i.e., 
$r_{i} > q_{i} \geq s_{\rm alg}(r_{i};\sigma|S)$ or 
$r_{i} < q_{i} \leq s_{\rm alg}(r_{i};\sigma|S)$. 
\end{enumerate}
\end{definition}
\begin{definition} \label{def-faithful}
Let {\sc alg} be an online/offline algorithm for ${\rm OFAL}(k,\ell)$. 
We say that 
{\sc alg} is {\sf faithful} if 
for any request sequence $\sigma=r_{1}\cdots r_{k\ell}$ and 
any request sequence $\tau=q_{1}\cdots q_{k\ell}$ that is closer to $S$ than 
$\sigma$ w.r.t.~{\sc alg},
$s_{\rm alg}(r_{i};\sigma|S) = s_{\rm alg}(q_{i};\tau|S)$ 
for each $1 \leq i \leq k\ell$. 
\end{definition}

Then we have the following lemma on {\sc opt} for ${\rm OFAL}(k,\ell)$. 
\begin{lemma} \label{lemma-opt-faithful}
{\sc opt} is faithful for ${\rm OFAL}(k,\ell)$.
\end{lemma}
\noindent {\bf Proof:} 
We consider {\sc opt} for ${\rm OFAL}(k,\ell)$. 
For a request sequence $\sigma=r_{1}\cdots r_{k\ell}$, 
let $r_{i}$ be a request such that $\leng{r_{i}-s_{\rm opt}(r_{i};\sigma|S)}>0$. 
For the request $r_{i}$, let $r_{i}'$ be a request such that 
$r_{i} > r_{i}' \geq s_{\rm opt}(r_{i};\sigma|S)$ or 
$r_{i} < r_{i}' \leq s_{\rm opt}(r_{i};\sigma|S)$. 
Define a request 
sequence $\tau=q_{1}\cdots q_{k\ell}$ 
by replacing $r_{i}$ with $r_{i}'$ in $\sigma$, i.e., 
for each $1 \leq h \leq k\ell$, 
\[
q_{h}=\left\{
\begin{array}{cl}
r_{h} &  h \neq i;\\
r_{h}' &  h=i
\end{array} \right.
\]
Note that the request sequence $\tau$ is closer to $S$ than $\sigma$ w.r.t.~{\sc opt}. 
Let  
\begin{eqnarray*}
s_{\rm opt}(\sigma|S)
&=&(s_{\rm opt}(r_{1};\sigma|S),\ldots,s_{\rm opt}(r_{k\ell};\sigma|S));\\
s_{\rm opt}(\tau|S)
&=&(s_{\rm opt}(q_{1};\tau|S),\ldots,s_{\rm opt}(q_{k\ell};\tau|S)), 
\end{eqnarray*}
and assume that $s_{\rm opt}(\sigma|S)\neq s_{\rm opt}(\tau|S)$. 
Since $s_{\rm opt}(\sigma|S)$ and $s_{\rm opt}(\tau|S)$ are optimal matchings 
for $\sigma$ and $\tau$, respectively,  it is immediate that 
\begin{eqnarray*}
\sum_{h=1}^{k\ell} \leng{r_{h}-s_{\rm opt}(r_{h};\sigma|S)}
&  = & 
\sum_{h=1}^{k\ell} \leng{q_{h}-s_{\rm opt}(r_{h};\sigma|S)}
+\leng{q_{i}-r_{i}}\\
& \geq & \sum_{h=1}^{k\ell} \leng{q_{h}-s_{\rm opt}(q_{h};\sigma|S)} 
+\leng{q_{i}-r_{i}}\\
& \geq & \sum_{h=1}^{k\ell} \leng{r_{h} -s_{\rm opt}(q_{h};\sigma|S)} \geq 
\sum_{h=1}^{k\ell} \leng{r_{h}-s_{\rm opt}(r_{h};\sigma|S)}. 
\end{eqnarray*}
This implies that $\sum_{h} \leng{r_{h}-s_{\rm opt}(r_{h};\sigma|S)}=
\sum_{h} \leng{r_{h}-s_{\rm opt}(q_{h};\sigma|S)}$. 
Then it follows that 
$s_{\rm opt}(\tau|S)$ is a common optimal matching for $\sigma$ and $\tau$.  
By iterating this process, we can conclude that 
{\sc opt} is faithful 
for ${\rm OFAL}(k,\ell)$. \BQED\medskip

To analyze the competitive ratio for faithful algorithms, the following notion is 
useful. 
\begin{definition} \label{def-opposite}
Let {\sc alg} be an online algorithm for ${\rm OFAL}(k,\ell)$ 
and $\sigma=r_{1}\cdots r_{k\ell}$ be a request sequence. We say that 
$\sigma$ is {\sf opposite} w.r.t.~{\sc alg} if 
for each $1 \leq i \leq k\ell$, 
\[
r_{i} \in [s_{\rm alg}(r_{i};\sigma|S),s_{\rm opt}(r_{i};\sigma|S)] ~\vee~
r_{i} \in [s_{\rm opt}(r_{i};\sigma|S),s_{\rm alg}(r_{i};\sigma|S)]. 
\] 
\end{definition}
\noindent The following lemma holds for request sequences 
w.r.t.~a faithful {\sc alg} for ${\rm OFAL}(k,\ell)$. 

\begin{lemma} \label{lemma-opposite}
Let {\sc alg} be a faithful online algorithm for ${\rm OFAL}(k,\ell)$.  
Then for any request sequence $\sigma$, there exists an opposite 
$\tau$ w.r.t.~{\sc alg} 
such that $\msc{rate}(\sigma) \leq 
\msc{rate}(\tau)$, where 
\[
\msc{rate}(\sigma) = \left\{
\begin{array}{cl}
\frac{\msc{alg}(\sigma|S)}{\msc{opt}(\sigma|S)} & \mbox{\rm if }
\msc{opt}(\sigma|S)>0;\\
\infty & \mbox{\rm if }\msc{opt}(\sigma|S)=0, \msc{alg}(\sigma|S)>0;\\
1 & \mbox{\rm if }\msc{opt}(\sigma|S)=\msc{alg}(\sigma|S)=0.
\end{array} \right.  
\]
\end{lemma}
\noindent {\bf Proof:} If $\sigma$ is opposite w.r.t.~{\sc alg}, then 
it suffices to set $\tau=\sigma$. Then we assume that 
$\sigma=r_{1}\cdots r_{k\ell}$ is {\it not\/} 
opposite w.r.t.~{\sc alg}. In this case, there exists $1 \leq i \leq k\ell$ 
such that \vspace*{-0.15cm} 
\begin{enumerate}
\item[(1)] $r_{i} < \min\{s_{\rm alg}(r_{i};\sigma|S), 
s_{\rm opt}(r_{i};\sigma|S)\}$; \vspace*{-0.25cm} 
\item[(2)] $r_{i} > \max\{s_{\rm alg}(r_{i};\sigma|S), 
s_{\rm opt}(r_{i};\sigma|S)\}$. 
\end{enumerate}\vspace*{-0.1cm} 
For the case (1), let $s \in S$ be the server closest to 
$r_{i}$ among $s_{\rm alg}(r_{i};\sigma|S)$ and 
$s_{\rm opt}(r_{i};\sigma|S)$, i.e., $s=\min\{s_{\rm alg}(r_{i};\sigma|S), 
s_{\rm opt}(r_{i};\sigma|S)\}$. 
Let $r_{i}'$ be a request that is located on $s$ 
and we define a request sequence $\tau=q_{1}\cdots q_{k\ell}$ as follows: 
for each $1 \leq h \leq k\ell$, 
\[
q_{h}= \left\{
\begin{array}{cl}
r_{h} & h \neq i;\\
r_{i}' & h=i.
\end{array} \right. 
\]
Note that $\tau$ is closer to $S$ than $\sigma$ w.r.t.~{\sc alg} and {\sc opt}. 
Since {\sc opt} is faithful for ${\rm OFAL}(k,\ell)$ by 
Lemma \ref{lemma-opt-faithful} and {\sc alg} is faithful for ${\rm OFAL}(k,\ell)$, 
we have that $s_{\rm alg}(r_{h};\sigma|S)=s_{\rm alg}(q_{h};\tau|S)$ and 
$s_{\rm opt}(r_{h};\sigma|S)=s_{\rm opt}(q_{h};\tau|S)$ 
for each $1 \leq h \leq k\ell$. Thus it follows that 
\[
\msc{rate}(\sigma) 
= \frac{\msc{alg}(\sigma|S)}{\msc{opt}(\sigma|S)} = 
\frac{\msc{alg}(\tau|S)+\leng{r_{i}-s}}{\msc{opt}(\tau|S)+\leng{r_{i}-s}} \leq 
\frac{\msc{alg}(\tau|S)}{\msc{opt}(\tau|S)} 
= \msc{rate}(\tau). 
\]
For the case (2), 
the argument similar to that of the case (1) holds. 
Iterate this process until $\tau$ gets opposite w.r.t.~{\sc alg}, and 
this completes the proof of the lemma.  \BQED

%
\subsection{Faithful MPFS Algorithms}  \label{subsec-faithful-mpfs}
%
In this subsection, we introduce crucial notions of a {\it characteristic\/} permutation,  
a {\it single\/} tour, and {\it multiple} tours. These notions provide a general 
framework for the analysis of the competitive ratio 
for faithful algorithms in ${\cal MPFS}$. 

\begin{definition} \label{def-characteristic-permutation}
Let $S=\{s_{1},\ldots,s_{k}\}$ be the set of $k$ servers 
on a line.  For an online~algorithm {\sc alg} and 
a request sequence $\sigma=r_{1}\cdots r_{k}$, we say that 
a bijection $\pi_{\sigma}^{\rm alg}:S\to S$ 
is a {\sf characteristic permutation} for $\sigma$ w.r.t.~{\sc alg} 
if $\pi_{\sigma}^{\rm alg}: s_{\rm opt}(r_{i};\sigma|S) \mapsto 
s_{\rm alg}(r_{i};\sigma|S)$ for each $1 \leq i \leq k$. 
We say that a request sequence $\sigma$ 
has a {\sf single tour} 
$T_{\sigma}^{\rm alg}$ 
on $S$ w.r.t.~{\sc alg} 
if $\pi_{\sigma}^{\rm alg}$ 
is cyclic on $S$, and 
$\sigma$ has {\sf multiple tours} $\{T_{\sigma}^{{\rm alg},i}\}_{i=1}^{t}$ 
if $\pi_{\sigma}^{\rm alg}$ is not cyclic on $S$. 
\end{definition}

For a request sequence $\sigma$, assume that $\sigma$ has a 
single tour $T_{\sigma}^{\rm alg}:s_{i_{1}}\to \cdots \to s_{i_{k}} \to s_{i_{1}}$ 
on $S$ w.r.t.~{\sc alg}. Then we define 
the length of $T_{\sigma}^{\rm alg}$~by 
\[ 
\ell(T_{\sigma}^{\rm alg}) = \leng{s_{i_{k}}-s_{i_{1}}} + \sum_{j=1}^{k-1}
\leng{s_{i_{j+1}}-s_{i_{j}}}. 
\] 
\noindent For an opposite request sequence $\sigma$ w.r.t.~faithful {\sc alg}, 
the following properties hold: 
\begin{property} \label{property-opposite}
Let $\sigma=r_{1}\cdots r_{k}$ be an opposite request sequence 
w.r.t.~faithful {\sc alg} and assume that $\sigma$ has a 
single tour $T_{\sigma}^{\rm alg}: s_{i_{1}} \to 
\cdots \to s_{i_{k}} \to s_{i_{1}}$ 
on $S$. Then \vspace*{-0.15cm}
\begin{enumerate}
\item[(1)] for each $1 \leq j \leq k$, 
there exists a request $r$ (in $\sigma$) 
that is located between $s_{i_{j}}$~and~$s_{i_{j+1}}$, 
where we regard $s_{i_{k+1}}$ as $s_{i_{1}}$, i.e., \vspace*{-0.15cm}
\begin{enumerate}
\item[(a)] if $s_{i_{j}} < s_{i_{j+1}}$, then 
$r \in [s_{i_{j}},s_{i_{j+1}}]$\vspace*{-0.1cm}
\item[(b)] if $s_{i_{j+1}} < s_{i_{j}}$, then $r \in [s_{i_{j+1}},s_{i_{j}}]$, 
\end{enumerate}\vspace*{-0.15cm}
and has ${\rm type}_{\rm alg}(r) =\pair{s_{i_{j+1}},s_{i_{j}}}$. \vspace*{-0.15cm}
\item[(2)] $\msc{alg}(\sigma|S)+\msc{opt}(\sigma|S)=\ell(T_{\sigma}^{\rm alg})$. 
\end{enumerate}
\end{property}
\noindent To derive an upper bound on the competitive ratio of 
faithful {\sc alg} 
for ${\rm OFAL}(k,\ell)$, we deal with the case that a request sequence $\sigma$ 
has a single tour 
in Theorems \ref{thm-upper-grdy-1} and \ref{thm-upper-idas-1}, while 
we deal with the case that $\sigma$ 
has multiple tours 
in Theorems \ref{thm-upper-grdy-2} and \ref{thm-upper-idas-2}. 

To show that any faithful algorithm is $c$-competitive, 
the following lemma is crucial, especially for the proofs of 
Theorems \ref{thm-upper-grdy-1} and \ref{thm-upper-idas-1}. 
\begin{lemma} \label{lemma-framework}
Let {\sc alg} be faithful for ${\rm OFAL}(k,\ell)$ and 
assume that 
an opposite request sequence $\sigma$ 
has a single tour 
$T_{\sigma}^{\rm alg}$ on $S$ w.r.t.~{\sc alg}. 
If there exists a function $H(T_{\sigma}^{\rm alg}) \in \mathbb{R}$ 
such that $\msc{opt}(\sigma|S) \geq \frac{H(T_{\sigma}^{\rm alg})}{c+1}$ and 
$\ell(T_{\sigma}^{\rm alg}) \leq H(T_{\sigma}^{\rm alg})$, 
then $\msc{alg}(\sigma|S) 
\leq c \cdot \msc{opt}(\sigma|S)$. 
\end{lemma}
\noindent {\bf Proof:} From Property \ref{property-opposite}(2), we have
that $\msc{alg}(\sigma|S)+\msc{opt}(\sigma|S)=\ell(T_{\sigma}^{\rm alg})$. Thus 
\[
\msc{alg}(\sigma|S)+\msc{opt}(\sigma|S)=\ell(T_{\sigma}^{\rm alg}) \leq 
H(T_{\sigma}^{\rm alg}) \leq (c+1)\cdot \msc{opt}(\sigma|S), 
\]
and this implies that 
$\msc{alg}(\sigma|S)\leq c \cdot \msc{opt}(\sigma|S)$. \BQED\medskip

In the remainder of the paper, we will simply 
use $T_{\sigma}$, $\{T_{\sigma}^{i}\}_{i=1}^{t}$, and $\pi_{\sigma}$ instead of 
$T_{\sigma}^{\rm alg}$, $\{T_{\sigma}^{{\rm alg},i}\}_{i=1}^{t}$, and 
$\pi_{\sigma}^{\rm alg}$, respectively, 
when {\sc alg} is clear from the context. 
%
\section{Competitive Ratio of Greedy Algorithm} \label{sec-cr-grdy}
%
In this section, we define one of the most natural 
algorithms for ${\rm OFA}(k,\ell)$ that is referred to as a 
{\it greedy\/} algorithm \cite{KalP1995}, and discuss the basic properties 
of the greedy algorithm. 

Before introducing the greedy algorithm for ${\rm OFA}(k,\ell)$, we begin with 
presenting a notion of {\it consuming\/} pairs of a request sequence $\sigma$. 
\begin{definition} \label{def-consuming-pair}
Let {\sc alg} be a faithful algorithm for ${\rm OFAL}(k,\ell)$ and 
$\sigma$ be an opposite request sequence w.r.t.~{\sc alg} with 
a single tour $T_{\sigma}: s_{h_{1}}\to \cdots \to 
s_{h_{k}}\to s_{h_{1}}$ on $S$. 
We say that a pair $(r_{i},r_{j})$ of requests is 
{\sf consuming} in $\sigma$ 
if $(s_{h_{p}},s_{h_{q}})$ is conflicting in $T_{\sigma}$, 
where~$s_{h_{p}}=s_{\rm opt}(r_{i};\sigma|S)$ and 
$s_{h_{q}}=s_{\rm opt}(r_{j};\sigma|S)$. 
Let ${\rm cs}_{\rm alg}(\sigma)$ be the set of all consuming 
pairs in $\sigma$,~i.e., 
${\rm cs}_{\rm alg}(\sigma)=\{(r_{i},r_{j}): (r_{i},r_{j}) \mbox{ is consuming 
in $\sigma$}\}$, and 
$f_{\rm bij}: {\rm cf}(T_{\sigma}) \to {\rm cs}_{\rm alg}(\sigma)$ be a bijection 
that maps $(s_{h_{p}},s_{h_{q}})$ to $(r_{i},r_{j})$ as above. 
\end{definition}

%
\begin{remark} \label{rem-cfcs}
%
From Definition \ref{def-consuming-pair}, it is immediate that 
$\leng{{\rm cf}(T_{\sigma})}=\leng{{\rm cs}_{\rm alg}(\sigma)}$. 
\end{remark}

Informally, we say that an algorithm for ${\rm OFA}(k,\ell)$ is {\it greedy} 
if the current request  is matched with the nearest free server. More formally, 
we have the following definition. 
\begin{definition} \label{def-grdy}
Let {\sc alg} be an online algorithm for ${\rm OFA}(k,\ell)$ and 
$\sigma =r_{1}\cdots r_{i} \cdots r_{n}$ be a request sequence 
such that $n=k\ell$. We say that 
{\sc alg} is a {\sf greedy} algorithm (denoted by {\sc grdy}), 
if {\sc alg} matches a request $r_{i}$ with 
the nearest\footnote{~If there exist at least two nearest free servers 
for the request $r_{i}$, then {\sc grdy} chooses  
the one with the {\it largest\/} index as the matching server for $r_{i}$.} free 
server $s \in S$ for each $1 \leq i \leq n$. 
\end{definition}
\noindent Kalyanasundaram and Pruhs \cite{KalP1995} 
showed that {\sc grdy} is $(2^{k}-1)$-competitive 
for ${\rm OFA}(k,1)$ 
and Kalyanasundaram and Pruhs \cite{KalP1998} 
mentioned that {\sc grdy} is also $(2^{k}-1)$-competitive 
for ${\rm OFA}(k,\ell)$ without proof. 
Since $\msc{grdy} \in {\cal MPFS}$ 
(see Definitions \ref{def-mpfs} and \ref{def-grdy}), 
Corollary \ref{cor-mpfs-separable} immediately provides a formal proof 
of the following result: 
\begin{corollary} \label{cor-gredy-ofa}
For any $\ell \geq 1$, {\sc grdy} is $(2^{k}-1)$-competitive for 
${\rm OFA}(k,\ell)$. 
\end{corollary}

The following lemma is essential for the subsequent discussions on {\sc grdy}. 
\begin{lemma} \label{lemma-grdy-faithful}
{\sc grdy} is faithful for ${\rm OFAL}(k,\ell)$. 
\end{lemma}
\noindent {\bf Proof:} 
Consider {\sc grdy} for ${\rm OFAL}(k,\ell)$. 
Let $\sigma=r_{1}\cdots r_{k\ell}$ and 
$\tau=q_{1}\cdots q_{k\ell}$ be 
request sequences, where $\tau$ is closer to $S$ than $\sigma$ w.r.t.~{\sc grdy}. 
For each $1 \leq i \leq k\ell$, {\sc grdy} 
matches a request $r_{i}$ with the server $s_{\rm grdy}(r_{i};\sigma|S)$. 
Since $s_{\rm grdy}(r_{i};\sigma|S)$ is the closest free server 
to both $r_{i}$ and $q_{i}$ for each $1 \leq i \leq k\ell$, 
we have that 
$s_{\rm grdy}(r_{i};\sigma|S)=s_{\rm grdy}(q_{i};\tau|S)$ 
for each $1 \leq i \leq k\ell$, i.e., {\sc grdy} is faithful 
for ${\rm OFAL}(k,\ell)$. \BQED\medskip%

Ahmed et al.~\cite{ARK2020} showed 
that {\sc grdy} is $4k$-competitive for ${\rm OFAL}_{eq}(k,\ell)$ with 
an informal proof. 
In this section, we 
show that ${\cal R}(\msc{grdy})=4k-5$ 
for ${\rm OFAL}_{eq}(k,\ell)$. In fact, we show that 
${\cal R}(\msc{grdy}) \geq 4k-5$ in Theorem \ref{thm-lower-grdy} 
and ${\cal R}(\msc{grdy}) \leq 4k-5$ in Corollary \ref{cor-upper-grdy-eq}, 
which generalizes the result by Itoh et al.~\cite{IMS2021}, i.e., 
${\cal R}(\msc{grdy})=3=4\cdot 2-5$ 
for ${\rm OFAL}_{eq}(2,\ell)$. 
\begin{remark} \label{rem-robust}
Since {\sc robust-matching} \cite{R2016} 
matches a request $r_{i}$ with a server depending on the  
positions of requests $r_{1},\ldots,r_{i}$ observed so far, 
we have that 
$\msc{robust-matching} \not \in {\cal MPFS}$. Thus, although 
{\sc robust-matching} for ${\rm OFAL}(k,1)$ is 
$O(\log k)$-competitive \cite{R2016,R2018}, this 
cannot be applied to ${\rm OFAL}(k,\ell)$ for $\ell>1$. \QED
\end{remark}

In the following subsections, we analyze the competitive ratio of {\sc grdy} for 
${\rm OFAL}(k,\ell)$. More precisely, we derive a lower bound 
on the competitive ratio of {\sc grdy} for ${\rm OFAL}_{eq}(k,\ell)$ 
in Section \ref{subsec-lower-grdyl} and upper bounds 
on the competitive ratio of {\sc grdy} for ${\rm OFAL}(k,\ell)$ 
and ${\rm OFAL}_{eq}(k,\ell)$  in Section \ref{subsec-upper-grdyl}. 
%
\subsection{A Lower Bound for the Competitive Ratio} 
\label{subsec-lower-grdyl}
%
In this subsection, we construct an {\it adversarial\/} request sequence $\sigma$ 
to derive a lower bound on 
the competitive ratio of  {\sc grdy} for ${\rm OFAL}_{eq}(k,\ell)$ with 
$k \geq 2$. 
\begin{theorem} \label{thm-lower-grdy}
For ${\rm OFAL}_{eq}(k,\ell)$ with $k \geq 2$, 
${\cal R}(\msc{grdy}) \geq 4k-5$. 
\end{theorem}
\noindent {\bf Proof:} For simplicity, assume that 
$s_{j}=j-1$ for each $1 \leq j \leq k$. We construct a request sequence 
$\sigma=\sigma_{1}\cdots \sigma_{\ell}$ such that $\msc{grdy}(\sigma|S) =
(4k-5)\cdot \msc{opt}(\sigma|S)$, where $\sigma_{i} = r_{1}^{i}\cdots r_{k}^{i}$. 
For each $1 \leq i \leq \ell$, 
let $r_{1}^{i}=\frac{1}{2}$ and 
$r_{j}^{i}=s_{j}=j-1$ for each $2 \leq j \leq k$. 
By Definition \ref{def-grdy}, 
%
%
\begin{eqnarray*}
\msc{grdy}(\sigma|S) & = & \left\{\frac{1}{2} + (k-2)+(k-1)\right\}\cdot \ell 
= \frac{4k-5}{2}\cdot \ell;\\
\msc{opt}(\sigma|S) & = & \frac{1}{2}\cdot \ell. 
\end{eqnarray*}
Thus for the request sequence $\sigma$ defined above, 
it follows that %
\[ 
\frac{\msc{grdy}(\sigma|S)}{\msc{opt}(\sigma|S)} = 
\frac{\frac{4k-5}{2}\cdot \ell}{\frac{1}{2}\cdot \ell} = 4k-5
\]
and this implies that ${\cal R}(\msc{grdy}) \geq 4k-5$. \BQED

%
\subsection{An Upper Bound for the Competitive Ratio}
\label{subsec-upper-grdyl}
%
In this subsection, we investigate the properties of {\sc grdy}  
and derive an upper bound on the competitive ratio of {\sc grdy} 
for ${\rm OFAL}(k,\ell)$, which leads to the {\it matching\/} upper bound 
for ${\rm OFAL}_{eq}(k,\ell)$. 
From Corollary \ref{cor-mpfs-separable} and 
the fact that $\msc{grdy} \in {\cal MPFS}$ 
for ${\rm OFAL}(k,\ell)$, 
it suffices to analyze the competitive ratio of {\sc grdy} for ${\rm OFAL}(k,1)$. 
In this subsection,~we 
consider only request sequences $\sigma$ of length $k$. 
For $k=1$, let $U(S)=1$ and for $k \geq 2$, let 
\[ 
U(S)=\frac{s_{k}-s_{1}}{d_{min}}
=\frac{\sum_{j=1}^{k-1}d_{j}}{\min_{1 \leq  j \leq k-1}d_{j}}. 
\] 

As shown in Lemma \ref{lemma-grdy-faithful}, we already know that 
{\sc grdy} is faithful for ${\rm OFAL}(k,\ell)$. Then from Lemma 
\ref{lemma-opposite}, it suffices to consider 
opposite request sequences w.r.t.~{\sc grdy} to derive an upper bound 
on the competitive ratio of {\sc grdy}.
In the remainder of this subsection, 
we assume that a request sequence $\sigma$ is opposite 
w.r.t.~{\sc grdy}. 
%
\subsubsection{Single Tour for {\footnotesize GRDY}}
\label{subsubsec-single-grdy}
%
We first consider the case that a request sequence $\sigma$ 
has a single tour w.r.t.~{\sc grdy} 
(and in Section \ref{subsubsec-multiple-grdy}, 
we also consider the case that $\sigma$ has 
multiple tours w.r.t.~{\sc grdy}). 
\begin{lemma} \label{lemma-consuming-pair}
$\msc{opt}(\sigma|S)\geq \frac{d_{min}}{2}\cdot  
\leng{{\rm cs}_{\rm grdy}(\sigma)}$ for any opposite 
request sequence $\sigma$. 
\end{lemma}
\noindent {\bf Proof:} Fix an opposite request sequence $\sigma$ 
w.r.t.~{\sc grdy} arbitrarily and we 
partition ${\rm cs}_{\rm grdy}(\sigma)$ into 
${\rm cs}_{\rm grdy}^{+}(\sigma)$ and ${\rm cs}_{\rm grdy}^{-}(\sigma)$ 
as follows: 
\begin{eqnarray*}
{\rm cs}_{\rm grdy}^{+}(\sigma) & = & \{(r_{i},r_{j}): 
(r_{i},r_{j}) \in {\rm cs}_{\rm grdy}(\sigma) \wedge i<j\};\\
{\rm cs}_{\rm grdy}^{-}(\sigma) & = & \{(r_{i},r_{j}): 
(r_{i},r_{j}) \in {\rm cs}_{\rm grdy}(\sigma) \wedge i>j\}. 
\end{eqnarray*}
For each request $a$ in 
$\sigma$, let ${\rm cs}_{\rm grdy}^{+}(a;\sigma)$ be 
the set of consuming pairs (in $\sigma$) 
of the form $(a,\ast) \in {\rm cs}_{\rm grdy}^{+}(\sigma)$, i.e., 
${\rm cs}_{\rm grdy}^{+}(a;\sigma)=
\{(a,r):(a,r) \in {\rm cs}_{\rm grdy}^{+}(\sigma)\}$, and  
enumerate nonempty ${\rm cs}_{\rm grdy}^{+}(a;\sigma)$'s by 
${\rm cs}_{\rm grdy}^{+}(a_{1};\sigma), \ldots, 
{\rm cs}_{\rm grdy}^{+}(a_{\mu};\sigma)$. 
Note that  
${\rm cs}_{\rm grdy}^{+}(a_{1};\sigma), \ldots, 
{\rm cs}_{\rm grdy}^{+}(a_{\mu};\sigma)$ is a 
partition of ${\rm cs}_{\rm grdy}^{+}(\sigma)$. 
For each request $b$ in 
$\sigma$, let ${\rm cs}_{\rm grdy}^{-}(b;\sigma)$ be 
the set of consuming pairs (in $\sigma$) of the form 
$(\ast,b) \in {\rm cs}_{\rm grdy}^{-}(\sigma)$, i.e., 
${\rm cs}_{\rm grdy}^{-}(b;\sigma)=\{(r,b):(r,b) \in 
{\rm cs}_{\rm grdy}^{-}(\sigma)\}$, and 
in a way similar to the definition of ${\rm cs}_{\rm grdy}^{+}(a;\sigma)$'s, 
we use  
${\rm cs}_{\rm grdy}^{-}(b_{1};\sigma), \ldots, 
{\rm cs}_{\rm grdy}^{-}(b_{\nu};\sigma)$ to denote a partition of 
${\rm cs}_{\rm grdy}^{-}(\sigma)$. It is immediate that 
\[
\leng{{\rm cs}_{\rm grdy}(\sigma)} = 
\leng{{\rm cs}_{\rm grdy}^{+}(\sigma)} + \leng{{\rm cs}_{\rm grdy}^{-}(\sigma)} = 
\sum_{i=1}^{\mu} \leng{{\rm cs}_{\rm grdy}^{+}(a_{i};\sigma)} +
\sum_{j=1}^{\nu} \leng{{\rm cs}_{\rm grdy}^{-}(b_{j};\sigma)}. 
\]
Note that $\{a_{1},\ldots,a_{\mu}\} \cap \{b_{1},\ldots,b_{\nu}\}= \emptyset$. 
If $\msc{opt}(a_{i};\sigma|S) \geq \frac{d_{min}}{2}\cdot 
\leng{{\rm cs}_{\rm grdy}^{+}(a_{i};\sigma)}$ for each $1 \leq i \leq \mu$ and 
$\msc{opt}(b_{j};\sigma|S) \geq \frac{d_{min}}{2}\cdot 
\leng{{\rm cs}_{\rm grdy}^{-}(b_{j};\sigma)}$ for each $1 \leq j \leq \nu$, then 
it follows that 
\begin{eqnarray*}
\lefteqn{\msc{opt}(\sigma|S) \geq 
\sum_{i=1}^{\mu} \msc{opt}(a_{i};\sigma|S) + 
\sum_{j=1}^{\nu} \msc{opt}(b_{j};\sigma|S)}\\
& \geq & 
\sum_{i=1}^{\mu} \frac{d_{min}}{2} \cdot 
\leng{{\rm cs}_{\rm grdy}^{+}(a_{i};\sigma)} + 
\sum_{j=1}^{\nu} \frac{d_{min}}{2} \cdot 
\leng{{\rm cs}_{\rm grdy}^{-}(b_{j};\sigma)}
= \frac{d_{min}}{2}\cdot \leng{{\rm cs}_{\rm grdy}(\sigma)}. 
\end{eqnarray*}
Thus it suffices to show that (1) 
$\msc{opt}(a_{i};\sigma|S) \geq \frac{d_{min}}{2}\cdot 
\leng{{\rm cs}_{\rm grdy}^{+}(a_{i};\sigma)}$ for each $1 \leq i \leq \mu$~and 
(2) $\msc{opt}(b_{j};\sigma|S) \geq \frac{d_{min}}{2}\cdot 
\leng{{\rm cs}_{\rm grdy}^{-}(b_{j};\sigma)}$ for each $1 \leq j \leq \nu$. 

For the case (1), fix $1 \leq i \leq \mu$ arbitrarily and 
consider $\msc{opt}(a_{i};\sigma|S)$. 
We assume that ${\rm cs}_{\rm grdy}^{+}(a_{i};\sigma)=\{(a_{i},r_{i_{1}}),\ldots,
(a_{i},r_{i_{u}})\}$, where $i_{1},\ldots,i_{u}$ are ordered in such a way that 
\[
s_{\rm opt}(a_{i};\sigma|S) \leq s_{\rm grdy}(r_{i_{u}};\sigma|S) < \cdots < 
s_{\rm grdy}(r_{i_{1}};\sigma|S) < s_{\rm grdy}(a_{i};\sigma|S). 
\]
Since $a_{i}$ is the earliest request among $a_{i}, r_{i_{1}},\ldots,r_{i_{u}}$ 
by the definition of ${\rm cs}_{\rm grdy}^{+}(\sigma)$, 
we have that  
$s_{\rm grdy}(a_{i};\sigma|S), 
s_{\rm grdy}(r_{i_{1}};\sigma|S), \ldots, s_{\rm grdy}(r_{i_{u}};\sigma|S)$ are 
free just before {\sc grdy} matches $a_{i}$ with 
$s_{\rm grdy}(a_{i};\sigma|S)$. This implies that 
$a_{i} \geq 
\frac{s_{\rm grdy}(a_{i};\sigma|S)+s_{\rm grdy}(r_{i_{1}};\sigma|S)}{2}$. Then 
\begin{eqnarray*}
\lefteqn{\msc{opt}(a_{i};\sigma|S)=\leng{a_{i}-s_{\rm opt}(a_{i};\sigma|S)} 
= a_{i}-s_{\rm opt}(a_{i};\sigma|S)\geq a_{i}-s_{\rm grdy}(r_{i_{u}};\sigma|S)}\\
& = & 
a_{i}-s_{\rm grdy}(r_{i_{1}};\sigma|S)
+ s_{\rm grdy}(r_{i_{1}};\sigma|S) -
s_{\rm grdy}(r_{i_{u}};\sigma|S)\\
& \geq & \frac{s_{\rm grdy}(a_{i};\sigma|S)+s_{\rm grdy}(r_{i_{1}};\sigma|S)}{2}
-s_{\rm grdy}(r_{i_{1}};\sigma|S)
+ s_{\rm grdy}(r_{i_{1}};\sigma|S) -
s_{\rm grdy}(r_{i_{u}};\sigma|S)\\
& = & \frac{s_{\rm grdy}(a_{i};\sigma|S)-s_{\rm grdy}(r_{i_{1}};\sigma|S)}{2} 
+ s_{\rm grdy}(r_{i_{1}};\sigma|S) -
s_{\rm grdy}(r_{i_{u}};\sigma|S)\\
& \geq & \frac{d_{min}}{2} + \sum_{j=1}^{u-1} 
\{s_{\rm grdy}(r_{i_{j}};\sigma|S)-s_{\rm grdy}(r_{i_{j+1}};\sigma|S)\}\\
& \geq & \frac{d_{min}}{2}+
\sum_{j=1}^{u-1} d_{min} \geq \frac{d_{min}}{2}\cdot u = 
\frac{d_{min}}{2}\cdot \leng{{\rm cs}_{\rm grdy}^{+}(a_{i};\sigma)}, 
\end{eqnarray*}
where the 1st inequality is due to the assumption that 
$s_{\rm grdy}(r_{i_{u}};\sigma|S) \geq  s_{\rm opt}(a_{i};\sigma|S)$. 

For the case (2), 
fix $1 \leq j \leq \nu$ arbitrarily and 
consider $\msc{opt}(b_{j};\sigma|S)$. 
We assume that ${\rm cs}_{\rm grdy}^{-}(b_{j};\sigma)=\{(r_{j_{1}},b_{j}),\ldots,
(r_{j_{v}},b_{j})\}$, where $j_{1},\ldots,j_{v}$ are ordered in such a way that 
\[
s_{\rm grdy}(b_{i};\sigma|S) <
s_{\rm grdy}(r_{j_{1}};\sigma|S) < \cdots < 
s_{\rm grdy}(r_{j_{v}};\sigma|S) \leq  s_{\rm opt}(b_{j};\sigma|S). 
\]
In a way similar to the case (1), 
we can show that 
$\msc{opt}(b_{j};\sigma|S) \geq \frac{d_{min}}{2}\cdot 
\leng{{\rm cs}_{\rm grdy}^{-}(b_{j};\sigma)}$ 
for each $1 \leq j \leq \nu$, and this complete the proof the 
lemma. \BQED\medskip

By applying Lemma \ref{lemma-framework} to {\sc grdy} 
for ${\rm OFAL}(k,\ell)$, we can show the following theorem. 
\begin{theorem} \label{thm-upper-grdy-1}
For a request sequence $\sigma$, if 
$\sigma$ has a single tour w.r.t.~{\sc grdy} on $S$, 
then $\msc{grdy}(\sigma|S)\leq (4\cdot U(S)-1)\cdot \msc{opt}(\sigma|S)$. 
\end{theorem}
\noindent {\bf Proof:} For $k=1$, it is immediate that 
$\msc{grdy}(\sigma|S) =\msc{opt}(\sigma|S)
\leq (4U(S)-1)\cdot \msc{opt}(\sigma|S)$ with 
$U(S)=1$. 
For any $k \geq 2$, let 
$T_{\sigma}:s_{h_{1}}\to \cdots \to s_{h_{k}}\to s_{h_{1}}$ be a 
single tour on $S$ and 
$H(T_{\sigma})=\leng{{\rm tp}(T_{\sigma})} 
\cdot (s_{k}-s_{1})$.~Then it follows that 
\begin{eqnarray*}
\msc{opt}(\sigma|S) & \geq & \frac{d_{min}}{2}\cdot 
\leng{{\rm cs}_{\rm grdy}(\sigma)} 
=\frac{d_{min}}{2}\cdot 
\leng{{\rm cf}(T_{\sigma})}\\
& \geq &
\frac{d_{min}}{4}\cdot \leng{{\rm tp}(T_{\sigma})}
=\frac{\leng{{\rm tp}(T_{\sigma})}\cdot (s_{k}-s_{1})}{4\cdot U(S)}
=\frac{H(T_{\sigma})}{4 \cdot U(S)},
\end{eqnarray*}
where the 1st inequality follows from Lemma \ref{lemma-consuming-pair}, 
the 1st equality follows from Remark \ref{rem-cfcs}, and 
the 2nd inequality follows from Lemma \ref{lemma-conflicting}. 
It is easy to see that 
$\ell(T_{\sigma}) \leq H(T_{\sigma})$. Thus from Lemma 
\ref{lemma-framework}, we have that 
$\msc{grdy}(\sigma|S) \leq (4\cdot U(S)-1)\cdot \msc{opt}(\sigma|S)$. \BQED
%
\subsubsection{Multiple Tours for {\footnotesize GRDY}}
\label{subsubsec-multiple-grdy}
%
In general, all the opposite request sequences 
do not necessarily have a single tour w.r.t. {\sc grdy}. 
For the case that an opposite request sequence $\sigma$ has multiple tours 
$\{T_{\sigma}^{i}\}_{i=1}^{t}$ w.r.t. {\sc grdy}, we regard each $T_{\sigma}^{i}$ 
as a single tour for a subsequence $\sigma_{i}$ of $\sigma$ 
and derive an upper bound of the competitive ratio by combining each 
of them for the request sequence $\sigma$. 
\begin{theorem} \label{thm-upper-grdy-2}
${\cal R}(\msc{grdy}) \leq 4\cdot U(S)-1$ 
for ${\rm OFAL}(k,\ell)$. 
\end{theorem}
\noindent {\bf Proof:} In Theorem \ref{thm-upper-grdy-1}, 
we already showed that $\msc{grdy}(\sigma|S) \leq 
(4\cdot U(S)-1)\cdot \msc{opt}(\sigma|S)$ for any request sequence 
$\sigma=r_{1}\cdots r_{k}$ with  
a single tour $T_{\sigma}$ on $S$, i.e., a bijection $\pi_{\sigma}: 
s_{\rm opt}(r_{i};\sigma|S)\mapsto s_{\rm grdy}(r_{i};\sigma|S)$ is cyclic on $S$. 
In the remainder of the proof, we show that 
$\msc{grdy}(\sigma|S) \leq 
(4\cdot U(S)-1)\cdot \msc{opt}(\sigma|S)$ for any request sequence $\sigma$ 
with multiple tours  
$T_{\sigma}^{1},\ldots,T_{\sigma}^{t}$ on $S$, i.e., the bijection 
$\pi_{\sigma}$ is not cyclic on $S$. 
Assume that $\pi_{\sigma}=\pi_{\sigma}^{1} \circ \cdots \circ \pi_{\sigma}^{t}$ 
for some $t \geq 2$, where 
$\pi_{\sigma}^{h}$ is a cyclic permutation on $S_{h}$ and 
can be regarded as a directed cycle on $S_{h}$ 
for each $1 \leq h \leq t$. 
Note that $S_{1}, \ldots, S_{t}$ is a partition of $S$. 
For each $1 \leq h \leq t$, 
we define a subsequence  
$\sigma_{h} = r_{1}^{h}\cdots r_{k_{h}}^{h}$ of a request sequence 
$\sigma$ such that 
$s_{\rm grdy}(r_{j}^{h};\sigma_{h}|S) \in S_{h}$ and 
$s_{\rm opt}(r_{j}^{h};\sigma_{h}|S) \in S_{h}$  
for each $ 1 \leq j \leq k_{h}$. 
Then we have that $\leng{\sigma_{h}}=\leng{S_{h}}$ 
for each $1 \leq h \leq t$. The following claims hold. 
\begin{claim} \label{claim-grdy}
For each $1 \leq h \leq t$, 
$\msc{grdy}(\sigma_{h};\sigma|S)=\msc{grdy}(\sigma_{h};\sigma_{h}|S_{h})$. 
\end{claim}
\begin{claim} \label{claim-opt}
For each $1 \leq h \leq t$, 
$\msc{opt}(\sigma_{h};\sigma|S)=\msc{opt}(\sigma_{h};\sigma_{h}|S_{h})$. 
\end{claim}
\noindent The proofs of Claims \ref{claim-grdy} and \ref{claim-opt} 
are given in Sections \ref{subsec-claim-grdy} and \ref{subsec-claim-opt}, 
respectively.  
Recall that $\pi_{\sigma}^{h}$ is a cyclic permutation on $S_{h}$ 
for each $1\leq h \leq t$. Then   
from Theorem 
\ref{thm-upper-grdy-1}, we have that $\msc{grdy}(\sigma_{h};\sigma_{h}|S_{h}) 
\leq (4\cdot U(S_{h})-1)\cdot \msc{opt}(\sigma_{h};\sigma_{h}|S_{h})$ 
for each $1\leq h \leq t$. Thus 
\begin{eqnarray*}
\lefteqn{\msc{grdy}(\sigma|S) = \msc{grdy}(\sigma;\sigma|S) = 
\sum_{h=1}^{t} \msc{grdy}(\sigma_{h};\sigma|S)}\\
& = & \sum_{h=1}^{t} \msc{grdy}(\sigma_{h};\sigma_{h}|S_{h}) 
\leq \sum_{h=1}^{t} (4\cdot U(S_{h})-1)\cdot 
\msc{opt}(\sigma_{h};\sigma_{h}|S_{h})\\
& \leq & (4\cdot U(S)-1) \cdot \sum_{h=1}^{t}  
\msc{opt}(\sigma_{h};\sigma_{h}|S_{h})
= (4\cdot U(S)-1) \cdot \sum_{h=1}^{t}  \msc{opt}(\sigma_{h};\sigma|S)\\
& = & (4\cdot U(S)-1) \cdot  \msc{opt}(\sigma;\sigma|S) = 
(4\cdot U(S)-1) \cdot  \msc{opt}(\sigma|S), 
\end{eqnarray*}
where the 3rd equality is due to Claim \ref{claim-grdy}, 
the 1st inequality is due to Theorem \ref{thm-upper-grdy-1}, 
and the 4th equality is due to  Claim \ref{claim-opt}, 
and this completes the proof of the theorem. \BQED
%
\subsection{Competitive Ratio for Greedy Algorithm 
for {\boldmath ${\bf OFAL}_{eq}(k,\ell)$}} \label{subsec-ul-grdy}
%
As an immediate consequence, we have the following corollary to Theorems 
\ref{thm-lower-grdy} and \ref{thm-upper-grdy-2}. 
\begin{corollary} \label{cor-upper-grdy-eq}
For ${\rm OFAL}_{eq}(k,\ell)$ such that 
$k \geq 2$, ${\cal R}(\msc{grdy}) = 4k-5$. 
\end{corollary}
\noindent {\bf Proof:} Since the distance between adjacent 
servers $s_{j}$ and $s_{j+1}$ is the same, i.e., $d_{j}=s_{j+1}-s_{j}=d$ 
for each $1 \leq j \leq k-1$, it is immediate that 
\[
U(S)=\frac{s_{k}-s_{1}}{d_{min}}
=\frac{\sum_{j=1}^{k-1} \{s_{j+1}-s_{j}\}}{\min_{1\leq j \leq k-1} 
\{s_{j+1}-s_{j}\}}
=\frac{\sum_{j=1}^{k-1}d}{\min_{1 \leq  j \leq k-1}d} = k-1.
\] 
From Theorem \ref{thm-upper-grdy-2}, it follows that 
${\cal R}(\msc{grdy}) \leq 4(k-1)-1=4k-5$. Since 
${\cal R}(\msc{grdy})\geq 4k-5$ for ${\rm OFAL}_{eq}(k,\ell)$ by Theorem 
\ref{thm-lower-grdy}, we have that ${\cal R}(\msc{grdy})=4k-5$. \BQED
%
\section{A Lower Bound on the Competitive Ratio of MPFS}
\label{sec-lower-mpfs}
%
In this section, we derive a lower bound on the competitive ratio 
of algorithms in ${\cal MPFS}$. 
\begin{theorem} \label{thm-lower-mpfs}
Let $\msc{alg} \in {\cal MPFS}$ for ${\rm OFAL}(k,\ell)$. Then 
${\cal R}(\msc{alg})\geq 2L(S)+1$, 
where $L(S)=0$ for $k=1$, and for any $k\geq 2$, 
\[
L(S)=\frac{s_{k}-s_{1}}{d_{max}} = \frac{s_{k}-s_{1}}{\max_{1 \leq j \leq k-1} d_{j}}. 
\]
Note that $d_{j}=s_{j+1}-s_{j}$ for each $1 \leq j \leq k-1$ as defined in 
(\ref{eq-interval}). 
\end{theorem}
Before presenting the proof of Theorem \ref{thm-lower-mpfs}, we introduce 
several notions, e.g., surrounding servers \cite{KN2004,AFT2018}, 
surrounding-oriented algorithms \cite{KN2004,AFT2018}, 
specification of algorithms (in Definition \ref{def-specification}), and 
feature points (in Definition \ref{def-feature})
and we also provide several technical lemmas related to those notions. 
\begin{definition} \label{def-surround}
Given a request $r$ for ${\rm OFAL}(k,\ell)$, 
the {\sf surrounding servers} for $r$ are $s^{L}$ and $s^{R}$, where 
$s^{L}$ is 
the closest free server to the left of $r$ (if any) and $s^{R}$ is 
the closest free server 
to the right of $r$ (if any). If $r=s$ for some $s \in S$ and $s$ is free, 
then the surrounding 
server of $r$ is only the server $s$. 
\end{definition}
\begin{definition} \label{def-surround-oriented}
Let {\sc alg} be an online algorithm for ${\rm OFAL}(k,\ell)$. We say that 
{\sc alg} is {\sf surrounding-oriented} for a request sequence 
$\sigma$ if it matches every request $r$ of $\sigma$ with one of the 
surrounding servers of $r$. 
We say that {\sc alg} is 
{\sf surrounding-oriented} if it is 
surrounding-oriented for every request sequence $\sigma$. 
\end{definition}

For surrounding-oriented algorithms, the following useful lemma 
\cite{AFT2018,IMS2021} is known. 

\begin{lemma} \label{lemma-surrounding-alg}
Let {\sc alg} be an online algorithm for ${\rm OFAL}(k,\ell)$. Then there exists 
a surroun\-ding-oriented algorithm $\msc{alg}'$ for ${\rm OFAL}(k,\ell)$ such that 
$\msc{alg}'(\sigma) \leq \msc{alg}(\sigma)$ for any $\sigma$. 
\end{lemma}

According to Lemma \ref{lemma-surrounding-alg}, it suffices to consider only 
$\msc{alg} \in {\cal MPFS}$ that is 
surround\-ing-oriented. To complete the proof of Theorem 
\ref{thm-lower-mpfs}, the following notions are necessary. 

\begin{definition} \label{def-specification}
Let $\msc{ alg} \in {\cal MPFS}$ for ${\rm OFAL}(k,\ell)$. 
For any pair of $1 \leq i < j \leq k$,~we say that {\sc alg} follows  
the specification $\msc{spec}(i,j)$ if {\sc alg} matches a request $r$ with 
the server $s_{i}$ when $s_{i}$ is free, 
$s_{i+1},\ldots,s_{j}$ are full, and $r$ occurs on $s_{j}$. 
\end{definition}

\begin{definition} \label{def-feature}
For any $k\geq 3$, let $\msc{alg}  \in {\cal MPFS}$ for ${\rm OFAL}(k,\ell)$. 
We say that $P_{\rm alg}=\{p_{1},\ldots,p_{m}\} \subseteq \{2,\ldots,k-1\}$ is a set of 
{\sf feature points} of {\sc alg} if it satisfies the following conditions: 
Let $p_{0}=1$ and for each $0 \leq i \leq m-1$, 
\[ 
p_{i+1}=\max \{j \in \{p_{i}+1,\ldots,k-1\}: 
\mbox{{\sc alg} follows $\msc{spec}(p_{i},j)$}\}. 
\]
%
For $\msc{alg} \in {\cal MPFS}$, 
if $p_{1}$ cannot be defined, 
then $P_{\rm alg}=\emptyset$.  
\end{definition}

By classifying  algorithms in ${\cal MPFS }$ 
due to Definition \ref{def-feature}, 
we show Theorem \ref{thm-lower-mpfs}.\bigskip%

\noindent {\bf Proof of Theorem \ref{thm-lower-mpfs}:} 
For simplicity, assume that $s_{1}=0$ and then $L(S)=\frac{s_{k}}{d_{max}}$.~Let 
$a_{j}=\frac{s_{j+1}\cdot s_{k}}{s_{k}+d_{j}}$
for each $ 1 \leq j \leq k-1$, and it is immediate that 
$s_{j}<a_{j}<s_{j+1}$. 

Let $k=1$ and $\msc{alg}' \in {\cal MPFS}$ for ${\rm OFAL}(1,\ell)$. 
Since ${\cal R}(\msc{alg}')=1$ and $L(S)=0$ for $k=1$, we have that 
${\cal R}(\msc{alg}')=1 =2L(S)+1$. Let $k=2$ and 
$\msc{alg}'' \in {\cal MPFS}$ for ${\rm OFAL}(2,\ell)$. 
Since ${\cal R}(\msc{alg}'')\geq 3$ \cite[Theorem 3.7]{IMS2020} and 
$L(S)=1$ for $k=2$, we have that 
${\cal R}(\msc{alg}'')\geq 3 =2L(S)+1$. Thus 
it suffices to consider the case that $k\geq 3$. 

For $k \geq 3$, fix $\msc{alg}\in {\cal MPFS}$ for ${\rm OFAL}(k,\ell)$ 
arbitrarily  and let $P_{\rm alg}=\{p_{1},\ldots,p_{m}\} \subseteq \{2,\ldots,k-1\}$ 
be the set of feature points of {\sc alg}. 
To derive a lower bound~on~the~competitive ratio of 
$\msc{alg} \in {\cal MPFS}$, we construct a request sequence 
$\sigma$ as follows: \vspace*{-0.15cm} 
\begin{enumerate}
\item For each $1 \leq j \leq k$, generate $\ell-1$ requests on $s_{j}$, 
which leads to the state that the remaining capacity of  
$s_{j}$ is 1 for each $1 \leq j \leq k$. \vspace*{-0.15cm} 
\item For each $2 \leq j \leq p_{m}$ such that $j \not \in P_{\rm alg}$,
generate a request on $s_{j}$, which leads to 
the state that $s_{j}$ is full for each 
$2 \leq j \leq p_{m}$ such that $j \not \in P_{\rm alg}$, and 
the remaining capacity of $s_{h}$ is 1 for $h=1$, $h \in P_{\rm alg}$, or 
$p_{m}<h \leq k$. \vspace*{-0.15cm} 
\item Generate a request $r^{(1)}$ on $a_{p_{m}}$ and 
let $s^{(1)}$ be the server with which {\sc alg} matches $r^{(1)}$. 
Note that $s_{p_{m}} < a_{p_{m}} < s_{p_{m}+1}$. \vspace*{-0.15cm} 
\item Generate a request $r^{(i+1)}$ on $s^{(i)}$ for each $i \geq 1$,
and continue the process 
until a request is generated on $s_{1}$ or $s_{k}$. 
\end{enumerate}\vspace*{-0.15cm} 
Since {\sc alg} is surrounding-oriented, it suffices to 
consider the following two cases: 
(Case 1) {\sc alg} matches $r^{(1)}$ with $s_{p_{m}+1}$; 
(Case 2) {\sc alg} matches $r^{(1)}$ with $s_{p_{m}}$.  

{\sf (Case 1)} Since  
{\sc alg} matches $r^{(1)}$ with $s_{p_{m}+1}$, we have 
two surrounding servers $s_{p_{m}}$ and $s_{p_{m}+2}$ for  
$r^{(2)}$ appearing on $s_{p_{m}+1}$. If {\sc alg} matches $r^{(2)}$ with 
$s_{p_{m}}$, then this implies~that~{\sc alg} follows 
$\msc{spec}(p_{m},p_{m}+1)$, but $p_{m}$ is the last element of 
$P_{\rm alg}$. Thus 
{\sc alg} must match $r^{(2)}$~with $s_{p_{m}+2}$. 
From this observation, 
it is obvious that {\sc alg} matches $r^{(i+1)}$ with $s_{p_{m}+i+1}$ 
for each $1 \leq i \leq k-p_{m}-1$ and matches $r^{(k-p_{m}+1)}$ with  
$s_{p_{m}}$, where $r^{(k-p_{m}+1)}$ is the last request of 
the request sequence $\sigma$. Then we have that 
\begin{eqnarray*}
\msc{alg}(\sigma|S) & = &  
\sum_{i=1}^{k-p_{m}-1} (s_{p_{m}+i+1}-s_{p_{m}+i})
+ (s_{p_{m}+1}-a_{p_{m}}) + (s_{k}-s_{p_{m}})\\
& = & 2\cdot ( s_{k}-s_{p_{m}} ) - (a_{p_{m}}-s_{p_{m}}), 
\end{eqnarray*}
and it is immediate that $\msc{opt}(\sigma|S) \leq a_{p_{m}}-s_{p_{m}}$. 
Thus it follows that 
\begin{eqnarray*}
\frac{\msc{alg}(\sigma|S)}{\msc{opt}(\sigma|S)} & \geq & 
\frac{2\cdot (s_{k}-s_{p_{m}}) - (a_{p_{m}}-s_{p_{m}})}{a_{p_{m}}-s_{p_{m}}} = 
2 \cdot \frac{s_{k}-s_{p_{m}}}{a_{p_{m}}-s_{p_{m}}}-1\\
& = & 2 \cdot \frac{s_{k}-s_{p_{m}}}
{\frac{s_{p_{m}+1}\cdot s_{k}}{s_{k}+d_{p_{m}}}-s_{p_{m}}}-1
= 2 \cdot \frac{(s_{k}-s_{p_{m}})\cdot (s_{k}+d_{p_{m}})}
{d_{p_{m}}\cdot (s_{k}-s_{p_{m}})}-1\\
& = & 2\cdot \frac{s_{k}}{d_{p_{m}}}+1 \geq 
2 \cdot \frac{s_{k}}{d_{max}}+1 
= 2\cdot L(S)+1. 
\end{eqnarray*}
%

{\sf (Case 2)} Recalling that 
{\sc alg} matches $r^{(1)}$ with $s_{p_{m}}$, we have 
two surrounding servers $s_{p_{m-1}}$ and $s_{p_{m}+1}$ for 
$r^{(2)}$ appearing on $s_{p_{m}}$. 
Since {\sc alg} follows 
$\msc{spec}(p_{m-1},p_{m})$, 
{\sc alg} must match $r^{(2)}$ with $s_{p_{m-1}}$. 
From this observation, 
it is obvious that {\sc alg} matches $r^{(i+1)}$ with $s_{p_{m-i}}$ 
for each $1 \leq i \leq m$, where $p_{0}=1$, 
and matches $r^{(m+2)}$ with  
$s_{p_{m}+1}$, where $r^{(m+2)}$ is the last request of 
the request sequence $\sigma$. Then 
we have that 
\begin{eqnarray*}
\msc{alg}(\sigma|S) & = &  (a_{p_{m}}-s_{p_{m}}) +
(s_{p_{m}+1}-s_{1}) +\sum_{i=1}^{m} (s_{p_{m-i+1}}-s_{p_{m-i}})\\
& = & (s_{p_{m}+1}-s_{1})+
(a_{p_{m}}-s_{1}) = s_{p_{m}+1}+a_{p_{m}}, 
\end{eqnarray*}
where the last equality follows from 
the assumption that $s_{1}=0$, 
and it is immediate that $\msc{opt}(\sigma|S) \leq s_{p_{m}+1}-a_{p_{m}}$. 
Then it follows that 
\begin{eqnarray*}
\frac{\msc{alg}(\sigma|S)}{\msc{opt}(\sigma|S)} & \geq & 
\frac{s_{p_{m}+1}+a_{p_{m}}}{s_{p_{m}+1}-a_{p_{m}}} 
= \frac{s_{p_{m}+1}+\frac{s_{p_{m}+1}\cdot s_{k}}{s_{k}+d_{p_{m}}}}
{s_{p_{m}+1}-\frac{s_{p_{m}+1}\cdot s_{k}}{s_{k}+d_{p_{m}}}}
= \frac{2s_{p_{m}+1}\cdot s_{k}
+d_{p_{m}}\cdot s_{p_{m}+1}}{d_{p_{m}}\cdot s_{p_{m}+1}}\\
& = & 2 \cdot \frac{s_{k}}{d_{p_{m}}} +1 \geq 
2 \cdot \frac{s_{k}}{d_{max}} +1
= 2\cdot L(S)+1, 
\end{eqnarray*}
and this completes the proof of the theorem. \BQED\bigskip%

As an immediate corollary to Theorem \ref{thm-lower-mpfs}, 
we have the following lower bound on the competitive ratio of 
any $\msc{alg} \in {\cal MPFS}$ for ${\rm OFAL}_{eq}(k,\ell)$. 
\begin{corollary} \label{cor-lower-mpfs}
Let $\msc{alg} \in {\cal MPFS}$ for ${\rm OFAL}_{eq}(k,\ell)$. Then 
${\cal R}(\msc{alg})\geq 2k-1$. 
\end{corollary}
\noindent {\bf Proof:} Since $d_{j}=s_{j+1}-s_{j}=d$ for each $1 \leq j \leq k-1$, 
we have that 
\[
L(S)=\frac{s_{k}-s_{1}}{\max_{1 \leq j \leq k-1} d_{j}} = 
\frac{\sum_{j=1}^{k-1} (s_{j+1}-s_{j})}
{\max_{1 \leq j \leq k-1} (s_{j+1}-s_{j})} = 
\frac{(k-1)\cdot d}{d}=k-1. 
\]
Thus the corollary follows from 
Theorem \ref{thm-lower-mpfs}. \BQED
%
\section{An Optimal MPFS Algorithm} \label{sec-opt-mpfs}
%
In this section, we propose a new algorithm 
$\msc{idas} \in {\cal MPFS}$ 
(Interior Division for Adjacent Servers), and we show that 
{\sc idas} is $(2k-1)$-competitive for ${\rm OFAL}_{eq}(k,\ell)$. From 
Corollary \ref{cor-lower-mpfs}, we can conclude that {\sc idas} is best possible 
in the class ${\cal MPFS}$ for ${\rm OFAL}_{eq}(k,\ell)$. 
%
\subsection{A New Algorithm: Interior Division for Adjacency Servers} 
\label{subsec-idas}
%
Before presenting the algorithm {\sc idas}, we provide several notations. 
Fix $a,b \in \mathbb{R}$~with~$a<b$ arbitrarily. 
For any $x,y \in \mathbb{R}$ such that 
$a \leq x < y\leq b$, let $B(x,y)$ 
be the point that internally divides the line segment $[x,y]$ into 
$b-x$ to $y-a$, i.e., 
\[
B(x,y) = \frac{(b-x)y+(y-a)x}{(b-x)+(y-a)}=\frac{by-ax}{b-a+y-x}.
\]
Note that $x < B(x,y) < y$ and $B(x,y)$ implies a {\it boundary\/} between 
$x$ and $y$. 

Given the set $S=\{s_{1},\ldots,s_{k}\}$ of $k$ servers with 
$s_{1} < \cdots < s_{k}$, we fix parameters~$a,b \in \mathbb{R}$ such that 
$a \leq s_{1}<s_{k}\leq b$. 
Then 
the algorithm $\msc{idas}_{[a,b]}$ can be described 
in Algorithm~\ref{alg-idas}.

\begin{algorithm}
\begin{flushleft}
\begin{minipage}{14.5cm}\vspace*{0.2cm}
For a request $r$, let $\msc{ss}(r)$ be the set of surrounding 
servers for $r$. \vspace*{-0.15cm}
\begin{enumerate}
\item If $\leng{\msc{ss}(r)}=1$, then let $\msc{ss}(r)=\{s_{*}\}$, where 
$s_{*} \in S$ is the unique surrounding\\
server for $r$, and match $r$ with $s_{*}$. \vspace*{-0.25cm}
\item If $\leng{\msc{ss}(r)}=2$, then 
let $s^{L}$ be the left surrounding server for $r$ and 
$s^{R}$ be the\\
right surrounding server for $r$. \vspace*{-0.15cm}
\begin{enumerate}
\item[(a)] If $r \leq B(s^{L},s^{R})$, then match 
$r$ with $s^{L}$;\vspace*{-0.1cm}
\item[(b)] If $B(s^{L},s^{R}) < r$, then match $r$ with $s^{R}$. 
\end{enumerate}
\end{enumerate}
\caption{$\msc{idas}_{[a,b]}$ (Interior Division for Adjacent Servers)\label{alg-idas}}
\end{minipage}
\end{flushleft}\vspace*{-0.5cm}
\end{algorithm}
\noindent It is immediate that $\msc{idas}_{[a,b]}$ is surrounding-oriented (see 
Definition \ref{def-surround-oriented}). 
In a way similar to Lemma \ref{lemma-grdy-faithful}, 
we can show the following lemma: 
\begin{lemma} \label{lemma-idas-faithful}
$\msc{idas}_{[a,b]}$ is faithful for ${\rm OFAL}(k,\ell)$. 
\end{lemma}
\noindent Hence, to derive an upper bound on the competitive 
ratio of $\msc{idas}_{[a,b]}$, 
it suffices to consider only 
opposite request sequences w.r.t.~$\msc{idas}_{[a,b]}$ (see 
Lemma \ref{lemma-opposite}). 

To observe that $\msc{idas}_{[a,b]} \in {\cal MPFS}$, 
the following property of $B(\ast,\ast)$ is crucial and the boundary $B(*,*)$ 
naturally induces a total order $\preceq_{\rho}$. 
\begin{property} \label{property-B}
Fix $a,b \in \mathbb{R}$~with~$a<b$ arbitrarily. 
For 
any $x,y,z\in \mathbb{R}$ such that~$a\leq x < y < z \leq b$, 
$B(x,y)<B(x,z)<B(y,z)$. 
\end{property}
\noindent {\bf Proof:} This follows from the straightforward calculations: 
\begin{eqnarray*}
B(x,z)-B(x,y) & = & \frac{(b-a)(z-y)(b-x)}{(b-a+z-x)(b-a+y-x)}>0;\\
B(y,z)-B(x,z) & = & \frac{(b-a)(y-x)(z-a)}{(b-z+y-a)(b-a+z-x)}>0, 
\end{eqnarray*}
where the inequalities follow from the assumption 
that $a\leq x<y<z \leq b$. \BQED\medskip%

We define the following binary relation $\preceq_{\rho}$ 
on $[a,b]$ with a parameter $\rho \in \mathbb{R}$. 
\begin{definition} \label{def-total-order}
For any $a,b \in \mathbb{R}$, let $[a,b]$ be the closed interval and fix 
$\rho \in \mathbb{R}$ arbitrarily. 
For any $x,y \in [a,b]$, we write $x \preceq_{\rho} y$ 
if one of the following conditions holds: (1) 
$x=y$; (2) $x < y$ and $B(x,y)<\rho$; (3) $y < x$ and $\rho \leq B(y,x)$. 
\end{definition}
For the binary relation $\preceq_{\rho}$ on $[a,b]$, 
the following result holds. 
\begin{theorem} \label{thm-total-order}
For any $\rho \in \mathbb{R}$, $\preceq_{\rho}$ is 
a total order on the closed interval $[a,b]$. 
\end{theorem}
\noindent The proof of the theorem is straightforward and is given in Appendix 
\ref{app-proof-total-order}. We summarize~the properties of the total order 
$\preceq_{\rho}$ in the following remark. 
\begin{remark} \label{rem-property-to}
For any $a,b \in \mathbb{R}$ such that $a<b$, let $[a,b]$ be the closed interval. 
Then for any $x,y \in [a,b]$ and any $\rho \in \mathbb{R}$, 
the following properties hold:\vspace*{-0.25cm} 
\begin{enumerate}
\item[(1)] if $\rho < x < y$, then 
$y \preceq_{\rho} x \preceq_{\rho} \rho$; \vspace*{-0.25cm} 
\item[(2)] if $x < y < \rho$, then $x \preceq_{\rho} y \preceq_{\rho} \rho$; \vspace*{-0.25cm} 
\item[(3)] if $\rho <a$, then 
$x \preceq_{\rho} y$ iff $x \geq y$; \vspace*{-0.25cm} 
\item[(4)] if $b<\rho$, then 
$x \preceq_{\rho} y$ iff $x \leq y$; \vspace*{-0.25cm} 
\item[(5)] if $\rho \in [a,b]$, then 
$x \preceq_{\rho} \rho$.\vspace*{-0.25cm} 
\end{enumerate}
The property (5) implies that $\rho$ is the maximum in $[a,b]$ w.r.t.~the 
total order $\preceq_{\rho}$. \QED
\end{remark}

From 
Remark~\ref{rem-property-to}, the following alternative definition is equivalent 
to that of $\msc{idas}_{[a,b]}$. 
\begin{definition} \label{def-idas}
For a request sequence $\sigma=r_{1}\cdots r_{i} \cdots r_{k\ell}$, 
the algorithm $\msc{idas}_{[a,b]}$ (Interior Division for Adjacent Servers) for 
${\rm OFAL}(k,\ell)$ works as follows: For each $1 \leq i \leq k\ell$, it 
matches a request $r_{i}$ 
with the highest free server\footnote{~For a request $r$, we say that 
$s \in S$ is the highest 
free server w.r.t.~$\preceq_{r}$ if $s$ is free and 
$s' \preceq_{r} s$ for all free servers 
$s' \in S$ just before matching $r$ to a server.} 
$s \in S$ w.r.t.~the total order $\preceq_{r_{i}}$. 
%
\end{definition}
From Definition \ref{def-idas}, it is immediate that 
$\msc{idas}_{[a,b]} \in {\cal MPFS}$. 
%
\subsection{An Upper Bound on the Competitive Ratio} 
\label{subsec-upper-idas}
%
In this subsection, 
we derive an upper bound on the competitive 
ratio of $\msc{idas} _{[a,b]}$ for ${\rm OFAL}(k,\ell)$, 
which leads to show that $\msc{idas} _{[a,b]}$ is best possible 
in ${\cal MPFS}$ for ${\rm OFAL}(k,\ell)$. 
%
\subsubsection{Single Tour for {\footnotesize IDAS}}
\label{subsubsec-single-idas}
%
Similarly to the discussion on {\sc grdy} in 
Section \ref{subsec-upper-grdyl}, 
we first consider a request sequence $\sigma$ 
with a single tour w.r.t.~$\msc{idas}_{[a,b]}$ 
(and consider a request sequence $\sigma$ with 
multiple tours w.r.t.~$\msc{idas}_{[a,b]}$ in Section \ref{subsubsec-multiple-idas}). 

For a conflicting pair $(v_{i},v_{j})$ in a tour 
$T:v_{1}\to \cdots \to v_{n} \to v_{1}$, 
we have that~$v_{i} \leq v_{j+1} < v_{i+1} \leq v_{j}$ by definition. 
Then the following cases are possible: 
(1) $v_{i}=v_{j+1}$~and $v_{i+1} < v_{j}$; 
(2) $v_{i}< v_{j+1}$ and $v_{i+1} = v_{j}$; 
(3) $v_{i} < v_{j+1}$ and $v_{i+1} < v_{j}$; 
(4) $v_{i}=v_{j+1}$~and $v_{i+1} =v_{j}$. For $n>2$, the case (4) never occurs, 
because the case (4) implies that~$v_{i} \to v_{i+1} \to v_{i}$ is a tour of 
length 2, but $T$ is a tour of length $n>2$.
For $n>2$,  let 
\[
c(v_{i},v_{j}|T) = \left\{
\begin{array}{ccl}
b-v_{i} & & \mbox{if $v_{i}=v_{j+1}$};\\
v_{j}-a & & \mbox{if $v_{i+1}=v_{j}$};\\ 
b-a & & \mbox{if $v_{i}<v_{j+1}$ and $v_{i+1}<v_{j}$}, 
\end{array} \right.
\]
and 
$c(v_{1},v_{2}|T)=\leng{v_{2}-v_{1}}$ for $n=2$. Define the cost of $T$ by 
\[
C(T) = \sum_{(v_{i},v_{j})\in {\rm cf}(T)} c(v_{i},v_{j}|T).
\]
In a way similar to $C(T)$, we can define $C(\tilde{T})$ for the contracted tour 
$\tilde{T}$ of $T$. 
\begin{lemma} \label{lemma-ell-2c}
For a tour $T$, let $\tilde{T}$ be the contracted tour of $T$. Then 
$\ell(\tilde{T}) \leq 2\cdot C(\tilde{T})$. 
\end{lemma}
\noindent {\bf Proof:} For a tour $T$, 
let $\tilde{T}$ 
be the contracted tour with $2m$ turning points. 
We show the lemma by induction on $m\geq 1$. 
For $m=1$, it is immediate that $\tilde{T}_{1}: t_{1} \to t_{2}\to t_{1}$ 
with $t_{1}<t_{2}$ has a 
single conflicting pair $(t_{1},t_{2})$. Then 
$\ell(\tilde{T}_{1})= 2 \cdot (t_{2}-t_{1})=2 \cdot C(\tilde{T}_{1})$. 

For any $m \geq 2$, assume that $\ell(\tilde{T}_{m-1}) \leq 
2 \cdot C(\tilde{T}_{m-1})$ for any $\tilde{T}_{m-1}$ with $2(m-1)$~turning 
points. We 
show that 
$\ell(\tilde{T}_{m}) \leq 2 \cdot C(\tilde{T}_{m})$ for 
any $\tilde{T}_{m}: t_{1}\to \cdots \to t_{2m}\to t_{1}$. 
Since~$m\geq 2$, there must exist a detour $D$ in 
$\tilde{T}_{m}$ by Lemma \ref{lemma-detour}. 
Without loss of generality, assume that 
$t_{1}\to t_{2}\to t_{3}\to t_{4}$ 
is a detour $D$ in $\tilde{T}_{m}$, where 
$t_{1} < t_{3} < t_{2} < t_{4}$ or $t_{1} > t_{3} > t_{2} > t_{4}$. 
Consider the case that 
$t_{1} < t_{3} < t_{2} < t_{4}$ (the other case 
can be discussed~analogously).  
For $\tilde{T}_{m-1}^{*}: t_{1}\to t_{4}\to \cdots \to 
t_{2m} \to t_{1}$ defined by 
contracting $D$ in $\tilde{T}_{m}$, we have 
that~$\ell(\tilde{T}_{m-1}^{*}) \leq 2 \cdot C(\tilde{T}_{m-1}^{*})$ by  
the induction hypothesis. 
Note that $t_{2}$ and $t_{3}$ are removed 
in $\tilde{T}_{m-1}^{*}$.~By~Def\-inition \ref{def-conflicting-tour}, 
$(t_{1},t_{j})$ is a conflicting pair in $\tilde{T}_{m-1}^{*}$ 
for each $4 \leq j \leq 2m$ 
if $t_{1} \leq t_{j+1} < t_{4} \leq t_{j}$. 
\begin{claim} \label{claim-1j}
For some $4 \leq j \leq 2m$, 
if $(t_{1},t_{j}) \in {\rm cf}(\tilde{T}_{m-1}^{*})$, then either 
$(t_{1},t_{j}) \in {\rm cf}(\tilde{T}_{m})$ or 
$(t_{3},t_{j}) \in {\rm cf}(\tilde{T}_{m})$ holds. 
\end{claim}
\noindent The proof of Claim \ref{claim-1j} is given  in Section 
\ref{subsec-claim-1j}. 
According to Claim \ref{claim-1j}, partition ${\rm cf}(\tilde{T}_{m-1}^{*})$ into 
${\rm cf}^{(1)}(\tilde{T}_{m-1}^{*}), {\rm cf}^{(3)}(\tilde{T}_{m-1}^{*})$, and  
${\rm cf}^{*}(\tilde{T}_{m-1}^{*})$ as follows: 
\begin{eqnarray*}
{\rm cf}^{(1)}(\tilde{T}_{m-1}^{*}) &= & 
\{(t_{1},t_{j}) \in {\rm cf}(\tilde{T}_{m-1}^{*}): 
(t_{1},t_{j}) \in {\rm cf}(\tilde{T}_{m})\};\\
{\rm cf}^{(3)}(\tilde{T}_{m-1}^{*}) & = & 
\{(t_{1},t_{j}) \in {\rm cf}(\tilde{T}_{m-1}^{*}): 
(t_{3},t_{j}) \in {\rm cf}(\tilde{T}_{m})\}
\setminus {\rm cf}^{(1)}(\tilde{T}_{m-1}^{*});\\
{\rm cf}^{*}(\tilde{T}_{m-1}^{*}) & = & 
\{(t_{i},t_{j}) \in {\rm cf}(\tilde{T}_{m-1}^{*}): i\neq 1\}, 
\end{eqnarray*}
and we also partition ${\rm cf}(\tilde{T}_{m})$ into 
${\rm cf}^{(2)}(\tilde{T}_{m}), {\rm cf}^{(3)}(\tilde{T}_{m}), 
{\rm cf}^{(1)}(\tilde{T}_{m})$, and ${\rm cf}^{*}(\tilde{T}_{m})$ as follows: 
\begin{eqnarray*}
{\rm cf}^{(2)}(\tilde{T}_{m}) & = & 
\{(t_{i},t_{j}) \in {\rm cf}(\tilde{T}_{m}): j=2\};\\
{\rm cf}^{(3)}(\tilde{T}_{m}) & = & 
\{(t_{i},t_{j}) \in {\rm cf}(\tilde{T}_{m}): j\neq 2 \wedge i=3\};\\
{\rm cf}^{(1)}(\tilde{T}_{m}) & = & 
\{(t_{i},t_{j}) \in {\rm cf}(\tilde{T}_{m}): j\neq 2 \wedge i=1\};\\
{\rm cf}^{*}(\tilde{T}_{m}) & = & 
\{(t_{i},t_{j}) \in {\rm cf}(\tilde{T}_{m}): j \neq 2 \wedge 
i \not \in \{1,3\}\}.
\end{eqnarray*}
For these partitions, we have the following claims: 
\begin{claim} \label{claim-ij*}
$c(t_{i},t_{j}|\tilde{T}_{m-1}^{*})
=c(t_{i},t_{j}|\tilde{T}_{m})$ for each 
$(t_{i},t_{j}) \in {\rm cf}^{*}(\tilde{T}_{m-1}^{*})$. 
\end{claim}
\begin{claim} \label{claim-1j3}
$c(t_{1},t_{j}|\tilde{T}_{m-1}^{*})
=c(t_{3},t_{j}|\tilde{T}_{m})$ for each 
$(t_{1},t_{j}) \in {\rm cf}^{(3)}(\tilde{T}_{m-1}^{*})$. 
\end{claim}
\begin{claim} \label{claim-1j1}
$c(t_{1},t_{j}|\tilde{T}_{m-1}^{*})
\leq c(t_{1},t_{j}|\tilde{T}_{m})$ for each 
$(t_{1},t_{j}) \in {\rm cf}^{(1)}(\tilde{T}_{m-1}^{*})$. 
\end{claim} 
\noindent The proofs of Claims \ref{claim-ij*}, \ref{claim-1j3}, 
and \ref{claim-1j1} are given in Sections 
\ref{subsec-claim-ij*}, \ref{subsec-claim-1j3}, 
and \ref{subsec-claim-1j1}, respectively. 
Then from these claims, it follows that 
\begin{eqnarray}
\lefteqn{C(\tilde{T}_{m-1}^{*}) = \sum_{(t_{i},t_{j}) \in {\rm cf}(\tilde{T}^{*}_{m-1})} 
c(t_{i},t_{j}|\tilde{T}^{*}_{m-1})}\nonumber\\
& = & \sum_{(t_{1},t_{j}) \in {\rm cf}^{(1)}(\tilde{T}_{m-1}^{*})}
c(t_{1},t_{j}|\tilde{T}_{m-1}^{*})
\nonumber\\& & ~~~~~~~ +  
\sum_{(t_{1},t_{j}) \in {\rm cf}^{(3)}(\tilde{T}_{m-1}^{*})}
c(t_{1},t_{j}|\tilde{T}_{m-1}^{*}) + 
\sum_{(t_{i},t_{j}) \in {\rm cf}^{*}(\tilde{T}_{m-1}^{*})}
c(t_{i},t_{j}|\tilde{T}_{m-1}^{*})\nonumber\\
& \leq & 
\sum_{(t_{1},t_{j}) \in {\rm cf}^{(1)}(\tilde{T}_{m-1}^{*})}
c(t_{1},t_{j}|\tilde{T}_{m}) 
\nonumber\\& & ~~~~~~~ +  
\sum_{(t_{1},t_{j}) \in {\rm cf}^{(3)}(\tilde{T}_{m-1}^{*})}
c(t_{3},t_{j}|\tilde{T}_{m})+  
\sum_{(t_{i},t_{j}) \in {\rm cf}^{*}(\tilde{T}_{m-1}^{*})}
c(t_{i},t_{j}|\tilde{T}_{m})\nonumber\\
& \leq & 
\sum_{(t_{i},t_{j}) \in {\rm cf}^{(1)}(\tilde{T}_{m})}
c(t_{i},t_{j}|\tilde{T}_{m}) +
\sum_{(t_{i},t_{j}) \in {\rm cf}^{(3)}(\tilde{T}_{m})}
c(t_{i},t_{j}|\tilde{T}_{m}) + 
\sum_{(t_{i},t_{j}) \in {\rm cf}^{*}(\tilde{T}_{m})}
c(t_{i},t_{j}|\tilde{T}_{m})\nonumber\\
& = & \sum_{(t_{i},t_{j}) \in {\rm cf}^{(1)}(\tilde{T}_{m})} c(t_{i},t_{j}|\tilde{T}_{m}) 
+ \sum_{(t_{i},t_{j}) \in {\rm cf}^{(2)}(\tilde{T}_{m})} 
c(t_{i},t_{j}|\tilde{T}_{m}) +
\sum_{(t_{i},t_{j}) \in {\rm cf}^{(3)}(\tilde{T}_{m})}
c(t_{i},t_{j}|\tilde{T}_{m}) \nonumber\\& & ~~~~~~~ +  
\sum_{(t_{i},t_{j}) \in {\rm cf}^{*}(\tilde{T}_{m})}
c(t_{i},t_{j}|\tilde{T}_{m})
- \sum_{(t_{i},t_{j}) \in {\rm cf}^{(2)}(\tilde{T}_{m})} c(t_{i},t_{j}|\tilde{T}_{m})\nonumber\\
& = & C(\tilde{T}_{m}) - 
\sum_{(t_{i},t_{j}) \in {\rm cf}^{(2)}(\tilde{T}_{m})}
c(t_{i},t_{j}|\tilde{T}_{m})\nonumber\\
& \leq & C(\tilde{T}_{m}) - c(t_{1},t_{2}|\tilde{T}_{m})
\leq C(\tilde{T}_{m}) - (t_{2}-t_{1}). \label{eq-ind-m}
\end{eqnarray}
Since $\ell(\tilde{T}_{m}) = \ell(\tilde{T}_{m-1}^{*})+2\cdot (t_{2}-t_{3})$, 
we have that 
\begin{eqnarray*}
\ell(\tilde{T}_{m}) & = & \ell(\tilde{T}_{m-1}^{*})+2\cdot (t_{2}-t_{3})
\leq 2\cdot C(\tilde{T}_{m-1}^{*})+2\cdot (t_{2}-t_{3})\\
& \leq & 2\cdot \{C(\tilde{T}_{m})-(t_{2}-t_{1})\}+2\cdot 
(t_{2}-t_{3})\\
& \leq & 2\cdot C(\tilde{T}_{m})+2\cdot (t_{1}-t_{3})\leq 
2\cdot C(\tilde{T}_{m}), 
\end{eqnarray*}
where the 1st inequality is due to the induction hypothesis, the 2nd 
inequality is due to Eq.(\ref{eq-ind-m}), and the last inequality is due to 
the assumption that 
$t_{1} < t_{3} < t_{2} < t_{4}$.\BQED
\begin{lemma} \label{lemma-CC}
For a tour $T$, let $\tilde{T}$ be the contracted tour of $T$. Then 
$C(\tilde{T}) \leq C(T)$. 
\end{lemma}
\noindent {\bf Proof:} For a tour $T:v_{1} \to \cdots \to v_{n} \to v_{1}$, 
let $\tilde{T}: t_{1}\to \cdots \to t_{2m}\to t_{1}$ be the contracted tour of $T$. 
Recall that the injection $f_{\rm inj}:{\rm cf}(\tilde{T}) \to {\rm cf}(T)$ 
was defined in Lemma~\ref{lemma-cf-T/contT}~as follows: For each $(t_{i},t_{j}) 
\in {\rm cf}(\tilde{T})$, we have that 
$t_{i} \leq t_{j+1}<t_{i+1}\leq t_{j}$~in~$\tilde{T}$.~Let~$T^{i}: t_{i}=v_{1}^{i} 
\to \cdots \to v_{x}^{i}=t_{i+1}$ (resp. $T^{j}: t_{j}=v_{1}^{j} \to \cdots \to 
v_{y}^{y}=t_{j+1}$) be the path from $t_{i}$ to $t_{i+1}$ 
(resp. from $t_{j}$ to $t_{j+1}$) in $T$. 
Let $1 \leq \alpha < x$ be the maximum 
with $v_{\alpha}^{i} \leq t_{j+1}$ and 
$1 \leq \beta < y$ be the maximum with 
$v_{\alpha}^{i} \leq t_{j+1} \leq v_{\beta+1}^{j} 
< v_{\alpha+1}^{i}  \leq v_{\beta}^{j}$,  
which implies that 
$(v_{\alpha}^{i},v_{\beta}^{j}) \in {\rm cf}(T)$. 
Then the injection $f_{\rm inj}: {\rm cf}(\tilde{T}) \to {\rm cf}(T)$ is given  
by $f_{\rm inj}: (t_{i},t_{j}) \mapsto (v_{\alpha}^{i},v_{\beta}^{j})$. 

We claim that 
$c(t_{i},t_{j}|\tilde{T}) \leq c(v_{\alpha}^{i},v_{\beta}^{j}|T)$ 
for each $(t_{i},t_{j}) \in {\rm cf}(\tilde{T})$. 
To show the claim, we consider the following four possible cases: 
(1) $c(v_{\alpha}^{i},v_{\beta}^{j}|T)=b-a$; 
(2) $c(v_{\alpha}^{i},v_{\beta}^{j}|T)=v_{\beta}^{j}-v_{\alpha}^{i}$; 
(3)~$c(v_{\alpha}^{i},v_{\beta}^{j}|T)=b-v_{\alpha}^{i}$; and 
(4) $c(v_{\alpha}^{i},v_{\beta}^{j}|T)=v_{\beta}^{j}-a$. 

For case (1), it is obvious that 
$c(t_{i},t_{j}|\tilde{T}) \leq b-a =c(v_{\alpha}^{i},v_{\beta}^{j}|T)$.
For case (2),  the tour $T$ must be 
a tour of length 2, i.e., 
$v_{\alpha}^{i} \to v_{\beta}^{j} \to v_{\alpha}^{i}$. Then the contracted tour of $T$ 
is given by $\tilde{T}: v_{\alpha}^{i} \to v_{\beta}^{j} \to v_{\alpha}^{i}$, where 
$t_{i}=v_{\alpha}^{i}$ and $t_{j}=v_{\beta}^{j}$. Thus it follows that 
\[
c(t_{i},t_{j}|\tilde{T})=t_{j}-t_{i}=v_{\beta}^{j}-v_{\alpha}^{i}
=c(v_{\alpha}^{i},v_{\beta}^{j}|T). 
\]
For case (3), we have that $v_{\beta+1}^{j}=v_{\alpha}^{i}$. 
If $v_{\beta+1}^{j} \in {\rm relay}(t_{j})$ or 
$v_{\alpha}^{i} \in {\rm relay}(t_{i})$, then it follows that  
$T$ visits $v_{\beta+1}^{j}$ or $v_{\alpha}^{i}$ more than once, 
but this is impossible by the definition 
of $T$.~Thus $v_{\beta+1}^{j} \not \in {\rm relay}(t_{j})$ and 
$v_{\alpha}^{i} \not \in {\rm relay}(t_{i})$.  
Since $1 \leq \alpha < x$ and $1 \leq \beta < y$,~we~have that 
$\alpha=1$ and $\beta+1=y$, i.e., 
$t_{i}=v_{1}^{i}=v_{\alpha}^{i}=v_{\beta+1}^{j}=v_{y}^{j}=t_{j+1}$. 
Then 
\[
c(t_{i},t_{j}|\tilde{T}) \leq b-t_{i} = b-v_{\alpha}^{i} = c(v_{\alpha}^{i},v_{\beta}^{j}|T).
\]
For case (4), it is immediate that $v_{\alpha+1}^{i}=v_{\beta}^{i}$. 
In a way similar to the case (3), 
we have that 
$\alpha+1=x$ and $\beta=1$, i.e., 
$t_{i+1}=v_{x}^{i}=v_{\alpha+1}^{i}=v_{\beta}^{j}=v_{1}^{j}=t_{j}$. 
Then 
\[
c(t_{i},t_{j}|\tilde{T}) \leq t_{j} -a= v_{\beta}^{j} -a= c(v_{\alpha}^{i},v_{\beta}^{j}|T).
\]
Thus we can conclude that 
$c(t_{i},t_{j}|\tilde{T}) \leq c(v_{\alpha}^{i},v_{\beta}^{j}|T)$ 
for each $(t_{i},t_{j}) \in {\rm cf}(\tilde{T})$. Then 
\begin{eqnarray*}
C(\tilde{T}) & = & \sum_{(t_{i},t_{j}) \in {\rm cf}(\tilde{T})} c(t_{i},t_{j}|\tilde{T})\\
& \leq & \sum_{(t_{i},t_{j}) \in {\rm cf}(\tilde{T})} c(f_{\rm inj}(t_{i},t_{j})|T) 
\leq \sum_{(v_{i},v_{j}) \in {\rm cf}(T)} c(v_{i},v_{j}|T) =C(T), 
\end{eqnarray*}
where the 2nd inequality is due to the fact that 
$\{f_{\rm inj}(t_{i},t_{j}): (t_{i},t_{j}) \in {\rm cf}(\tilde{T})\} 
\subseteq {\rm cf}(T)$. \BQED
\begin{lemma} \label{lemma-opt-C}
For any opposite request sequence $\sigma$, 
if 
$\sigma$ has a single tour $T_{\sigma}$ 
on $S$ w.r.t. $\msc{idas}_{[a,b]}$, then 
$\msc{opt}(\sigma|S) \geq \frac{d_{min}}{b-a+d_{min}}\cdot C(T_{\sigma})$. 
\end{lemma}
\noindent {\bf Proof:} Fix an opposite request sequence 
$\sigma=r_{1}\cdots r_{k}$ w.r.t.~$\msc{idas}_{[a,b]}$ arbitrarily, and let 
$T_{\sigma}: s_{h_{1}} \to \cdots \to s_{h_{k}} \to s_{h_{1}}$ be 
the single tour on $S$ w.r.t.~$\msc{idas}_{[a,b]}$.  
We partition the set ${\rm cs}_{\rm idas}(\sigma)$ of consuming pairs in $\sigma$ 
(see Definition \ref{def-consuming-pair}) 
into ${\rm cs}_{\rm idas}^{+}(\sigma)$  and ${\rm cs}_{\rm idas}^{-}(\sigma)$ 
as follows: 
\begin{eqnarray*}
{\rm cs}_{\rm idas}^{+}(\sigma) & = & 
\{(r_{i},r_{j}) \in {\rm cs}_{\rm idas}(\sigma): i<j\};\\
{\rm cs}_{\rm idas}^{-}(\sigma) & = & 
\{(r_{i},r_{j}) \in {\rm cs}_{\rm idas}(\sigma): i>j\}. 
\end{eqnarray*}
For each request $a$ in 
$\sigma$, let ${\rm cs}_{\rm idas}^{+}(a;\sigma)$ be 
the set of consuming pairs (in $\sigma$) 
of the form $(a,\ast) \in {\rm cs}_{\rm idas}^{+}(\sigma)$, i.e., 
${\rm cs}_{\rm idas}^{+}(a;\sigma)=
\{(a,r):(a,r) \in {\rm cs}_{\rm idas}^{+}(\sigma)\}$, and  
enumerate nonempty ${\rm cs}_{\rm idas}^{+}(a;\sigma)$'s by 
${\rm cs}_{\rm idas}^{+}(a_{1};\sigma), \ldots, 
{\rm cs}_{\rm idas}^{+}(a_{\mu};\sigma)$. 
Notice that  
${\rm cs}_{\rm idas}^{+}(a_{1};\sigma), \ldots, 
{\rm cs}_{\rm idas}^{+}(a_{\mu};\sigma)$ is a 
partition of ${\rm cs}_{\rm idas}^{+}(\sigma)$. 
For each request $b$ in 
$\sigma$, let ${\rm cs}_{\rm idas}^{-}(b;\sigma)$ be 
the set of consuming pairs (in $\sigma$) of the form 
$(\ast,b) \in {\rm cs}_{\rm idas}^{-}(\sigma)$, i.e., 
${\rm cs}_{\rm idas}^{-}(b;\sigma)=\{(r,b):(r,b) \in 
{\rm cs}_{\rm idas}^{-}(\sigma)\}$, and 
in a way similar to the definition of ${\rm cs}_{\rm idas}^{+}(a;\sigma)$'s, 
we use  
${\rm cs}_{\rm idas}^{-}(b_{1};\sigma), \ldots, 
{\rm cs}_{\rm idas}^{-}(b_{\nu};\sigma)$ to denote the partition of 
${\rm cs}_{\rm idas}^{-}(\sigma)$. 
As we mentioned in Remark \ref{rem-cfcs}, 
there exists~the~bijection~$f_{\rm bij}:{\rm cf}(T_{\sigma})\to {\rm cs}_{\rm idas}(\sigma)$. 
Then we have the following claims: 
\begin{claim} \label{claim-ai}
For each $1 \leq i \leq \mu$, the following inequality holds. 
\[
\msc{opt}(a_{i};\sigma|S) \geq \frac{d_{min}}{b-a+d_{min}} \ 
\sum_{(a_{i},r) \in {\rm cs}_{\rm idas}^{+}(a_{i};\sigma)} 
c\left(f_{\rm bij}^{-1}(a_{i},r)|T_{\sigma}\right).
\] 
\end{claim}
\begin{claim} \label{claim-bj}
For each $1 \leq j \leq \nu$, the following inequality holds. 
\[
\msc{opt}(b_{j};\sigma|S) \geq \frac{d_{min}}{b-a+d_{min}} \ 
\sum_{(r,b_{j}) \in {\rm cs}_{\rm idas}^{-}(b_{j};\sigma)} 
c\left(f_{\rm bij}^{-1}(r,b_{j})|T_{\sigma}\right).
\] 
\end{claim}
\noindent The proofs of Claims \ref{claim-ai} and \ref{claim-bj} are given in 
Sections \ref{subsec-claim-ai} and \ref{subsec-claim-bj}, respectively. Then 
\begin{eqnarray*}
C(T_{\sigma}) & = & \sum_{(s_{i},s_{j}) \in {\rm cf}(T_{\sigma})} 
c(s_{i},s_{j}|T_{\sigma}) = 
\sum_{(r_{i},r_{j}) \in {\rm cs}_{\rm idas}(\sigma)} 
c\left(f_{\rm bij}^{-1}(r_{i},r_{j})|T_{\sigma}\right)\\
& = & \sum_{(r_{i},r_{j}) \in {\rm cs}_{\rm idas}^{+}(\sigma)} 
c\left(f_{\rm bij}^{-1}(r_{i},r_{j})|T_{\sigma}\right)+ 
\sum_{(r_{i},r_{j}) \in {\rm cs}_{\rm idas}^{-}(\sigma)} 
c\left(f_{\rm bij}^{-1}(r_{i},r_{j})|T_{\sigma}\right)\\
& = & \sum_{i=1}^{\mu} \sum_{(a_{i},r) \in {\rm cs}_{\rm idas}^{+}(a_{i};\sigma)} 
c\left(f_{\rm bij}^{-1}(a_{i},r)|T_{\sigma}\right)+
\sum_{j=1}^{\nu} \sum_{(r,b_{j}) \in {\rm cs}_{\rm idas}^{-}(b_{j};\sigma)} 
c\left(f_{\rm bij}^{-1}(r,b_{j})|T_{\sigma}\right);\\
\msc{opt}(\sigma|S) & \geq & \sum_{i=1}^{\mu} \msc{opt}(a_{i};\sigma|S) + 
\sum_{j=1}^{\nu} \msc{opt}(b_{j};\sigma|S).
\end{eqnarray*}
Thus the lemma immediately follows from Claims \ref{claim-ai} 
and \ref{claim-bj}. \BQED\medskip

By applying Lemma \ref{lemma-framework} to $\msc{idas}_{[a,b]}$ 
for ${\rm OFAL}(k,\ell)$, we can show the following theorem. 
\begin{theorem} \label{thm-upper-idas-1}
Let $\sigma$ be an opposite request sequence w.r.t.~$\msc{idas}_{[a,b]}$ 
for ${\rm OFAL}(k,\ell)$. If $\sigma$ has a single tour on $S$, 
then 
\[
\msc{idas}_{[a,b]}(\sigma|S)\leq 
\left(2\cdot \frac{b-a}{d_{min}}+1\right)\cdot \msc{opt}(\sigma|S). 
\]
\end{theorem}
\noindent {\bf Proof:} Let $H(T_{\sigma})=2\cdot C(T_{\sigma})$. Then 
we have 
an upper bound on $\ell(T_{\sigma})$, i.e., 
\[ 
\ell(T_{\sigma}) = \ell(\tilde{T}_{\sigma}) \leq 2 \cdot C(\tilde{T}_{\sigma}) 
\leq 2 \cdot C(T_{\sigma})=H(T_{\sigma}), 
\]
where the 1st inequality follows from Lemma \ref{lemma-ell-2c} and 
the 2nd inequality follows from Lemma \ref{lemma-CC}. On the other hand, 
we also have a lower bound on $\msc{opt}(\sigma|S)$, i.e., 
\[ 
\msc{opt}(\sigma|S) \geq \frac{d_{min}}{b-a+d_{min}}\cdot C(T_{\sigma}) 
= \frac{2\cdot C(T_{\sigma})}{2\cdot \left(\frac{b-a}{d_{min}}+1\right)} = 
\frac{H(T_{\sigma})}{2 \cdot \left(\frac{b-a}{d_{min}}+1\right)}, 
\] 
where the inequality is due to Lemma \ref{lemma-opt-C}. 
Thus it follows that 
\[
\msc{idas}_{[a,b]}(\sigma|S) \leq \left(2 \cdot \frac{b-a}{d_{min}}+1\right) 
\cdot \msc{opt}(\sigma|S)
\]
by letting $c+1=2\cdot (\frac{b-a}{d_{min}}+1)$ in 
Lemma \ref{lemma-framework}. \BQED
%
%
\subsubsection{Multiple Tours for {\footnotesize IDAS}}
\label{subsubsec-multiple-idas}
%
In general, all the opposite request sequences 
do not necessarily have a single tour. 
Then we consider the case  
that an opposite request sequence $\sigma$ has multiple tours 
$T_{\sigma}^{1}, \ldots,T_{\sigma}^{t}$ 
%
\begin{theorem} \label{thm-upper-idas-2}
${\cal R}(\msc{idas}_{[a,b]})\leq 2\cdot \frac{b-a}{d_{min}}+1$ 
for ${\rm OFAL}(k,\ell)$. 
\end{theorem}
\noindent {\bf Proof:} In Theorem \ref{thm-upper-idas-1}, we already showed that 
$\msc{idas}_{[a,b]}(\sigma|S) \leq (2\cdot \frac{b-a}{d_{min}}+1)\cdot 
\msc{opt}(\sigma|S)$ 
for any request sequence $\sigma=r_{1}\cdots r_{k}$ with  
a single tour $T_{\sigma}$ on $S$, i.e., a bijection $\pi_{\sigma}: 
s_{\rm opt}(r_{i};\sigma|S)\mapsto s_{\rm idas}(r_{i};\sigma|S)$ is cyclic on $S$. 
In the remainder of the proof, we show that 
$\msc{idas}_{[a,b]}(\sigma|S) \leq 
(2\cdot \frac{b-a}{d_{min}}+1)\cdot 
\msc{opt}(\sigma|S)$ 
for any request sequence $\sigma$ 
with multiple tours  
$T_{\sigma}^{1},\ldots,T_{\sigma}^{t}$ on $S$, i.e., the bijection 
$\pi_{\sigma}$ is not cyclic on $S$. 
Assume that $\pi_{\sigma}=\pi_{\sigma}^{1} \circ \cdots \circ \pi_{\sigma}^{t}$ 
for some $t \geq 2$, where 
$\pi_{\sigma}^{h}$ is a cyclic permutation on $S_{h}$ and 
can be regarded as a directed cycle on $S_{h}$ 
for each $1 \leq h \leq t$. 
Note that $S_{1}, \ldots, S_{t}$ is a partition of $S$. 
For each $1 \leq h \leq t$, 
we define a subsequence  
$\sigma_{h} = r_{1}^{h}\cdots r_{k_{h}}^{h}$ of a request sequence 
$\sigma$ such that 
$s_{\rm idas}(r_{j}^{h};\sigma_{h}|S) \in S_{h}$ and 
$s_{\rm opt}(r_{j}^{h};\sigma_{h}|S) \in S_{h}$  
for each $ 1 \leq j \leq k_{h}$. 
Then we have that $\leng{\sigma_{h}}=\leng{S_{h}}$ 
for each $1 \leq h \leq t$. The following claims hold. 
\begin{claim} \label{claim-idas}
$\msc{idas}_{[a,b]}(\sigma_{h};\sigma|S)=
\msc{idas}_{[a,b]}(\sigma_{h};\sigma_{h}|S_{h})$ for each 
$1 \leq h \leq t$. 
\end{claim}
\noindent The proof of Claim \ref{claim-idas} is given in Section 
\ref{subsec-claim-idas}. 
Recall that $\pi_{\sigma}^{h}$ is cyclic on $S_{h}$ for each 
$1 \leq h \leq t$. Then from Theorem \ref{thm-upper-idas-1}, 
it follows that for each $1 \leq h \leq t$, 
\begin{equation}
\msc{idas}_{[a,b]}(\sigma_{h}; \sigma_{h}|S_{h}) \leq \left(
2 \cdot \frac{b-a}{d_{min}^{h}}+1\right)\cdot 
\msc{opt}(\sigma_{h};\sigma_{h}|S_{h}), 
\label{eq-app-1}
\end{equation}
where $d_{min}^{h}=
\min\{ \leng{s_{i}-s_{j}}: s_{i},s_{j} \in S_{h} \mbox{ $(i\neq j)$}\}$. 
Thus we have that 
\begin{eqnarray*}
\lefteqn{\msc{idas}_{[a,b]}(\sigma|S) = 
\msc{idas}_{[a,b]}(\sigma;\sigma|S) = 
\sum_{h=1}^{t} \msc{idas}_{[a,b]}(\sigma_{h};\sigma|S)}\\
& = & \sum_{h=1}^{t} \msc{idas}_{[a,b]}(\sigma_{h};\sigma_{h}|S_{h})
\leq \sum_{h=1}^{t} \left(2\cdot \frac{b-a}{d_{min}^{h}}+1\right)\cdot 
\msc{opt}(\sigma_{h};\sigma_{h}|S_{h})\\
& \leq & \left(2\cdot \frac{b-a}{d_{min}}+1\right)\cdot 
\sum_{h=1}^{t} \msc{opt}(\sigma_{h};\sigma_{h}|S_{h})\\
& = & \left(2\cdot \frac{b-a}{d_{min}}+1\right)\cdot 
\sum_{h=1}^{t} \msc{opt}(\sigma_{h};\sigma|S_{h})
= \left(2\cdot \frac{b-a}{d_{min}}+1\right)\cdot 
\msc{opt}(\sigma|S_{h}), 
\end{eqnarray*}
where the 3rd equality is due to Claim \ref{claim-idas}, the 1st inequality is due to 
(\ref{eq-app-1}), and the 4th equality is due to Claim \ref{claim-opt}, 
and this completes the proof of the theorem. \BQED\medskip

As an immediate consequence, we have the following corollary to 
Theorem \ref{thm-upper-idas-2}. 
\begin{corollary} \label{cor-idas-cr}
${\cal R}(\msc{idas}_{[s_{1},s_{k}]}) \leq 2k-1$ for 
${\rm OFAL}_{eq}(k,\ell)$.
\end{corollary}
\noindent {\bf Proof:} Apply Theorem \ref{thm-upper-idas-2} 
for ${\rm OFAL}_{eq}(k,\ell)$ by setting $a=s_{1}$ and $b=s_{k}$. Then 
\[
2 \cdot \frac{b-a}{d_{min}} +1= 2 \cdot \frac{s_{k}-s_{1}}{1}+1
= 2 \cdot \frac{k-1}{1}+1=2k-1. 
\]
Thus it follows 
${\cal R}(\msc{idas}_{[s_{1},s_{k}]}) \leq 2k-1$ for 
${\rm OFAL}_{eq}(k,\ell)$. \BQED
%
\section{Concluding Remarks} \label{sec-remark}
%
In this paper, we dealt with the online facility assignment problem 
${\rm OFA}(k,\ell)$, 
where $k\geq 1$ is the number of servers and $\ell\geq 1$ 
is a capacity for each server. As special cases of ${\rm OFA}(k,\ell)$, 
we also dealt with ${\rm OFA}(k,\ell)$ {\it on a line\/}, which is denoted by 
${\rm OFAL}(k,\ell)$ and ${\rm OFAL}_{eq}(k,\ell)$, where the latter 
is the case of 
${\rm OFAL}(k,\ell)$ with equidistant servers. 

In Section \ref{sec-mpfs}, we introduced the class of MPFS (Most 
Preferred Free Servers) algorithms and showed that any MPFS algorithm 
has the capacity-insensitive property (in Corollary \ref{cor-mpfs-separable}). 
In Section \ref{sec-faithful}, we formulated the {\it faithful\/} property 
crucial for the competitive analysis in the paper. 
In Section \ref{sec-cr-grdy}, we analyzed the competitive ratio of {\sc grdy} 
for ${\rm OFAL}_{eq}(k,\ell)$ and showed that ${\cal R}(\msc{grdy})=4k-5$   
(in Corollary \ref{cor-upper-grdy-eq}). 
In Section \ref{sec-lower-mpfs}, 
we showed that for ${\rm OFAL}_{eq}(k,\ell)$, 
${\cal R}(\msc{alg})\geq 2k-1$ for any $\msc{alg} \in {\cal MPFS}$ 
(in Corollary \ref{cor-lower-mpfs}). 
In Section \ref{sec-opt-mpfs}, 
we proposed a new MPFS algorithm {\sc idas} (Interior Division for Adjacent Servers) 
for ${\rm OFAL}(k,\ell)$ and showed that for ${\rm OFAL}_{eq}(k,\ell)$, 
${\cal R}(\msc{idas})\leq 2k-1$ (in Corollary \ref{cor-idas-cr}), i.e., 
{\sc idas} for ${\rm OFAL}_{eq}(k,\ell)$ is best possible 
in all of the MPFS algorithms. 

Notice that for ${\rm OFAL}_{eq}(k,\ell)$, 
any algorithm in ${\cal MPFS}$ has the 
capacity-insensitive property and the competitive ratio of 
$\msc{idas} \in {\cal MPFS}$ matches 
the lower bound of any algorithm in ${\cal MPFS}$. This implies that 
for ${\rm OFAL}_{eq}(k,\ell)$, 
there does not exist an algorithm in ${\cal MPFS}$ with the competitive ratio 
better than that of {\sc idas}. Thus  for ${\rm OFAL}_{eq}(k,\ell)$, 
one of the most interesting problems is to design 
capacity-insensitive algorithms not in ${\cal MPFS}$ with the better competitive 
ratio than that of {\sc idas}. 																																										
%

%
\appendix
%
\section{Proof of Claims in Section \ref{subsec-upper-grdyl}} 
\label{app-proof-claims-grdy}
%
\subsection{Proof of Claim \ref{claim-grdy}}
\label{subsec-claim-grdy}
%
For the set $S$ of $k$ servers and a request sequence $\sigma=r_{1}\cdots r_{k}$, 
consider the case that 
{\sc grdy} matches $r_{j}^{h}$ with  
$s_{j}^{h}=s_{\rm grdy}(r_{j}^{h};\sigma|S)$ for each $1 \leq j \leq k_{h}$. 
From the definition of {\sc grdy}, it is immediate that 
just before {\sc grdy} matches $r_{j}^{h}$ with $s_{j}^{h}$, all of 
$s_{j}^{h}.\ldots,s_{k_{h}}^{h}$ are free and $s_{j}^{h}$ is the nearest 
to $r_{j}^{h}$ among the free servers $s_{j}^{h}.\ldots,s_{k_{h}}^{h}$. 
This is preserved to the case that 
just before 
{\sc grdy} with the set $S_{h}$ of $k_{h}$ servers matches $r_{j}^{h}$ with 
a free server $s \in  S_{h}$ on a request sequence $\sigma_{h}$ as an input. 
Thus 
\[
s_{\rm grdy}(r_{j}^{h};\sigma|S) = s_{j}^{h}=
s_{\rm grdy}(r_{j}^{h};\sigma_{h}|S_{h})
\]
for each $1 \leq j \leq k_{h}$, and this completes the proof of 
the claim. 
%
\subsection{Proof of Claim \ref{claim-opt}}
\label{subsec-claim-opt}
%
For each $1 \leq h \leq t$, we have that 
\begin{equation}
\{s_{\rm opt}(r_{j}^{h};\sigma|S): 1 \leq j \leq k_{h}\}
=S_{h} = \{s_{\rm opt}(r_{j}^{h};\sigma_{h}|S_{h}): 1 \leq j \leq k_{h}\}, 
\label{eq-opt-servers}
\end{equation}
and $\msc{opt}(\sigma_{h};\sigma_{h}|S_{h}) \leq \msc{opt}(\sigma_{h};\sigma|S)$. 
Assume that there exists  a request sequence 
$\sigma_{g}$ such that 
$\msc{opt}(\sigma_{g};\sigma_{g}|S_{g}) < \msc{opt}(\sigma_{g};\sigma|S)$ 
and we use $\sigma-\sigma_{g}$ to denote the request sequence~defined 
by deleting $\sigma_{g}$ 
from $\sigma$. Then 
from (\ref{eq-opt-servers}), it is immediate that 
\begin{eqnarray*}
\msc{opt}(\sigma|S)=\msc{opt}(\sigma;\sigma|S) & =& 
\msc{opt}(\sigma_{g};\sigma|S)+\msc{opt}(\sigma-\sigma_{g};\sigma|S)\\
& > & 
\msc{opt}(\sigma_{g};\sigma_{g}|S_{g})
+\msc{opt}(\sigma-\sigma_{g};\sigma-\sigma_{g}|(S\setminus S_{g})), 
\end{eqnarray*}
and this contradicts the optimality of $\msc{opt}$ on $\sigma$. Thus 
for each $ 1 \leq h \leq t$, it follows that 
$\msc{opt}(\sigma_{h};\sigma|S)=\msc{opt}(\sigma_{h};\sigma_{h}|S_{h})$, 
and this completes the proof of the claim. 
%
\section{Proof of Theorem \ref{thm-total-order}} \label{app-proof-total-order}
%
For any $a,b \in \mathbb{R}$ such that $a<b$, let $[a,b]$ be the closed interval. 
Fix $\rho \in \mathbb{R}$ arbitrarily.\medskip\\
{\sf (Reflexivity)} For any $x \in [a,b]$, $x \preceq_{\rho} x$ 
by Definition \ref{def-total-order}. \smallskip\\
{\sf (Antisymmetry)} For any $x,y \in [a,b]$ such that $x \leq y$ 
(and the case that $y \geq x$ can be discussed analogously), 
assume that $x \preceq_{\rho} y$ and $y \preceq_{\rho} x$. 
Then we have that 
\begin{eqnarray*}
x \preceq_{\rho} y & \to & x=y \mbox{ or } B(x,y) < \rho;\\
y \preceq_{\rho} x & \to & y=x \mbox{ or } \rho \leq B(x,y).
\end{eqnarray*}
By the assumption that $x \preceq_{\rho} y$ and $y \preceq_{\rho} x$, 
the only possible case is $x=y$. \medskip\\
{\sf (Transitivity)} For any $x,y,z \in [a,b]$, assume that $x \preceq_{\rho} y$ and 
$y \preceq_{\rho} z$. If $x=y$ or $y=z$, then it is immediate that 
$x \preceq_{\rho} z$. We show that $x \preceq_{\rho} z$ 
for all the other cases. \vspace*{-0.15cm}
\begin{enumerate}
\item[(1)] $x < y < z$: It is immediate that $B(x,z)<B(y,z)$ 
by Property \ref{property-B} and  we also have that 
$B(y,z) < \rho$ by the assumption that $y \preceq_{\rho} z$. Then it follows that 
$x \preceq_{\rho} z$. \vspace*{-0.25cm}
\item[(2)] $x < z < y$: It is immediate that $B(x,z)<B(x,y)$ 
by Property \ref{property-B} and  we also have that 
$B(x,y) < \rho$ by the assumption that $x \preceq_{\rho} y $. 
Then it follows that $x \preceq_{\rho} z$. \vspace*{-0.25cm}
\item[(3)] $y < x < z$: It is obvious that $\rho \leq B(y,x)$ and $B(y,z) < \rho$ by 
the assumptions $x \preceq_{\rho} y$ and $y \preceq_{\rho} z$, 
respectively, and we also 
have that $B(y,x)<B(y,z)$ by Property \ref{property-B}. Then  
$\rho \leq B(y,z) < B(y,z) < \rho$, which is the contradiction. Thus for 
$y < x < z$, the assumptions that $x \preceq_{\rho} y$ and 
$y \preceq_{\rho} z$ do not hold. \vspace*{-0.25cm}
\item[(4)] $y < z < x$: We have that $\rho \leq B(y,x)$ by the assumption that 
$x \preceq_{\rho} y $ and $B(y,x)<B(z,x)$ by Property \ref{property-B}. 
Then it follows that $x \preceq_{\rho} z$. \vspace*{-0.25cm}
\item[(5)] $z < x < y$: It is obvious that $\rho \leq B(z,y)$ and $B(x,y) < \rho$ by 
the assumptions $y \preceq_{\rho} z$ and $x \preceq_{\rho} y$, 
respectively, and we also 
have that $B(z,y)<B(x,y)$ by Property \ref{property-B}. Then  
$\rho \leq B(z,y) < B(x,y) < \rho$, which is the contradiction. Thus for 
$z < x < y$, the assumptions that $x \preceq_{\rho} y$ and 
$y \preceq_{\rho} z$ do not hold. \vspace*{-0.25cm}
\item[(6)] $z < y < x$: We have that $\rho \leq B(z,y)$ by the assumption that 
$y \preceq_{\rho} z $ and $B(z,y)<B(z,x)$ by Property \ref{property-B}. 
Then it follows that $x \preceq_{\rho} z$. 
\end{enumerate}\vspace*{-0.15cm}
{\sf (Comparability)} For any $x,y \in [a,b]$, we show that $x \preceq_{\rho} y$ or 
$y \preceq_{\rho} x$. 

For the case that $x=y$, we have 
that $x \preceq_{\rho} y$ and $y \preceq_{\rho} x$. 
Consider the case that~$x < y$.  
From the definition of $\preceq_{\rho}$, it follows that if $\rho \leq B(x,y)$, 
then $y \preceq_{\rho} x$ and if $B(x,y) < \rho$, then $x \preceq_{\rho} y$. 
For the case that $y > x$, we can show that $x \preceq_{\rho} y$ 
or $y \preceq_{\rho} x$ analogously. 
%
\section{Proof of Claims in Section \ref{subsec-upper-idas}} 
\label{app-proof-claims}
%
\subsection{Proof of Claim \ref{claim-1j}}
\label{subsec-claim-1j}
%
Since $(t_{1},t_{j}) 
\in {\rm cf}(\tilde{T}_{m-1}^{*})$, we have that  
$t_{1}\leq t_{j+1}<t_{4}\leq t_{j}$ in $\tilde{T}_{m-1}^{*}$. 
If $t_{3} \leq t_{j+1}$ in $\tilde{T}_{m}$, 
then it is immediate that $t_{3}\leq t_{j+1}<t_{4}\leq t_{j}$, i.e., 
$(t_{3},t_{j}) \in {\rm cf}(\tilde{T}_{m})$. 
If $t_{3} > t_{j+1}$ in $\tilde{T}_{m}$, 
then from the assumption that 
$t_{1} < t_{3} < t_{2} < t_{4}$, it follows that 
$t_{1}\leq t_{j+1}<t_{3}<t_{2}<t_{4}\leq t_{j}$. 
%
%
%
%
This implies that $t_{1}\leq t_{j+1} < t_{2}< t_{j}$ in $\tilde{T}_{m}$, i.e., 
$(t_{1},t_{j}) \in {\rm cf}(\tilde{T}_{m})$. 
%
\subsection{Proof of Claim \ref{claim-ij*}}
\label{subsec-claim-ij*}
%
Since $(t_{i},t_{j}) \in {\rm cf}^{*}(\tilde{T}_{m-1}^{*})$ such that $i \neq 1$, 
we have that $t_{i} \leq t_{j+1} < t_{i+1}\leq t_{j}$, 
%
%
%
%
which is preserved in $\tilde{T}_{m}$. Thus it follows that 
$c(t_{i},t_{j}|\tilde{T}_{m-1}^{*})
= c(t_{i},t_{j}|\tilde{T}_{m})$. 
%
\subsection{Proof of Claim \ref{claim-1j3}}
\label{subsec-claim-1j3}
%
For a conflicting pair 
$(t_{1},t_{j}) \in {\rm cf}^{(3)}(\tilde{T}_{m-1}^{*})$, it follows that 
$(t_{3},t_{j}) \in {\rm cf}(\tilde{T}_{m})$ and 
$(t_{1},t_{j}) \not \in {\rm cf}(\tilde{T}_{m})$. Recall that 
$t_{1}<t_{3}<t_{2} < t_{4}$ for the detour $D$ in $\tilde{T}_{m}$ 
and $t_{3} \leq t_{j+1} < t_{4} \leq t_{j}$ for the conflicting pair 
$(t_{3},t_{j}) \in {\rm cf}(\tilde{T}_{m})$. Thus we have that 
$t_{1} < t_{3} \leq t_{j+1} < t_{4} \leq t_{j}$. 
%
%
%
%
Since 
$(t_{1},t_{j}) \not \in {\rm cf}(\tilde{T}_{m})$, we have that 
$t_{2} \leq t_{j+1}$, i.e., 
$t_{1} <t_{3}<t_{2} < t_{j+1} < t_{4} \leq t_{j}$. 
This implies that $t_{1}<t_{j+1}<t_{4}\leq t_{j}$ and 
$t_{3} <t_{j+1}<t_{4} \leq t_{j}$. 

If $t_{4}=t_{j}$, then we have that 
$c(t_{1},t_{4}|\tilde{T}_{m-1}^{*}) = t_{4}-a = 
c(t_{3},t_{4}|\tilde{T}_{m})$, 
and if $t_{4} < t_{j}$,~then~we have that
$c(t_{1},t_{j}|\tilde{T}_{m-1}^{*}) = b-a = 
c(t_{3},t_{j}|\tilde{T}_{m})$. 
%
\subsection{Proof of Claim \ref{claim-1j1}}
\label{subsec-claim-1j1}
%
Since $(t_{1},t_{j}) \in 
{\rm cf}^{(1)}(\tilde{T}_{m-1}^{*})$, we have that  
$t_{1} \leq t_{j+1} < t_{4} \leq t_{j}$ in $\tilde{T}_{m-1}^{*}$
and that $t_{1} \leq t_{j+1} < t_{2} \leq t_{j}$ in $\tilde{T}_{m}$. 
Let us consider the following two cases: $t_{1}<t_{j+1}$ and 
$t_{1} = t_{j+1}$. 

If $t_{1} < t_{j+1}$, then we have that 
$c(t_{1},t_{j}|\tilde{T}_{m-1}^{*}) \leq b-a = 
c(t_{1},t_{j}|\tilde{T}_{m})$.  
If $t_{1}=t_{j+1}$, then we consider the following cases: 
$t_{4}=t_{j}$ and $t_{4} < t_{j}$, and we have that 
\[
c(t_{1},t_{j}|\tilde{T}_{m-1}^{*}) = 
\left\{ \begin{array}{ccl}
t_{j}-t_{1} & & \mbox{if $t_{4}=t_{j}$};\\
b-t_{1} & & \mbox{if $t_{4}< t_{j}$}.
\end{array} \right.
\]
Thus it follows that 
$c(t_{1},t_{j}|\tilde{T}_{m-1}^{*})\leq b-t_{1} 
= c(t_{1},t_{j}|\tilde{T}_{m})$. 
%
\subsection{Proof of Claim \ref{claim-ai}} \label{subsec-claim-ai}
%
For each $1 \leq i \leq \mu$, assume that 
${\rm cs}_{\rm idas}^{+}(a_{i};\sigma) = \{(a_{i},r_{i_{1}}), \ldots,(a_{i},r_{i_{u}})\}$, 
where $i_{1},\ldots,i_{u}$ are ordered in such a way that 
\[
s_{\rm opt}(a_{i};\sigma|S) \leq 
s_{\rm idas}(r_{i_{u}};\sigma|S) < \cdots < 
s_{\rm idas}(r_{i_{2}};\sigma|S) < 
s_{\rm idas}(r_{i_{1}};\sigma|S) < 
s_{\rm idas}(a_{i};\sigma|S). 
\]
Since $a_{i}$ is the earliest request among $a_{i}, r_{i_{1}},\ldots,r_{i_{u}}$ 
by the definition of ${\rm cs}_{\rm idas}^{+}(\sigma)$, 
we have that  
$s_{\rm idas}(r_{i_{1}};\sigma|S)$ and $s_{\rm idas}(a_{i};\sigma|S)$ are free  
just before $a_{i}$ arrives. 
Let us consider 
the case that $s_{\rm opt}(a_{i};\sigma|S) < s_{\rm idas}(r_{i_{u}};\sigma|S)$. Then 
we have that 
\begin{eqnarray*}
\lefteqn{\msc{opt}(a_{i};\sigma|S) = a_{i}-s_{\rm opt}(a_{i};\sigma|S) \geq 
s_{\rm idas}(r_{i_{1}};\sigma|S) - s_{\rm opt}(a_{i};\sigma|S)}\\
& = &s_{\rm idas}(r_{i_{u}};\sigma|S)-s_{\rm opt}(a_{i};\sigma|S)+ 
\sum_{s=1}^{u-1} \left\{s_{\rm idas}(r_{i_{s}};\sigma|S)-
s_{\rm idas}(r_{i_{s+1}};\sigma|S)\right\}\\
& \geq & d_{min} + \sum_{s=1}^{u-1} d_{min} = u\cdot d_{min}
= \frac{d_{min}}{b-a+d_{min}}\cdot u \cdot (b-a+d_{min})\\
& \geq & \frac{d_{min}}{b-a+d_{min}} \cdot
\sum_{(a_{i},r) \in {\rm cs}_{\rm idas}^{+}(a_{i};\sigma)}
c\left(f_{\rm bij}^{-1}(a_{i},r)|T_{\sigma}\right). 
\end{eqnarray*}
We turn to consider 
the case that $s_{\rm opt}(a_{i};\sigma|S) = s_{\rm idas}(r_{i_{u}};\sigma|S)$. Then 
it is immediate that  
$c(s_{\rm opt}(a_{i};\sigma|S),s_{\rm opt}(r_{i_{u}};\sigma|S)|T_{\sigma}) 
\leq b-s_{\rm opt}(a_{i};\sigma|S)$. Note that 
\begin{eqnarray*}
\sum_{(a_{i},r) \in {\rm cs}_{\rm idas}^{+}(a_{i};\sigma)} 
c\left( f_{\rm bij}^{-1}(a_{i},r)|T_{\sigma}\right) & = & 
\sum_{s=1}^{u} c\left( f_{\rm bij}^{-1}(a_{i},r_{i_{s}})|T_{\sigma}\right)\\
& \leq &
b-s_{\rm opt}(a_{i};\sigma|S)+ 
\sum_{s=1}^{u-1} c\left( f_{\rm bij}^{-1}(a_{i},r_{i_{s}})|T_{\sigma}\right)\\
& \leq &
b-s_{\rm opt}(a_{i};\sigma|S)+\sum_{s=1}^{u-1} (b-a) \\
& = & b-s_{\rm opt}(a_{i};\sigma|S)+(u-1) \cdot (b-a). 
\end{eqnarray*}
From the definition of 
$\msc{idas}_{[a,b]}$, it is immediate that 
\begin{equation}
s_{\rm idas}(r_{i_{1}};\sigma|S) \leq
B(s_{\rm idas}(r_{i_{1}};\sigma|S), s_{\rm idas}(a_{i};\sigma|S))
\leq a_{i}. \label{eq-2nd}
\end{equation}
Then $\msc{opt}(a_{i};\sigma|S)$ can be estimated as follows: 
\begin{eqnarray*}
\lefteqn{\msc{opt}(a_{i};\sigma|S) = a_{i} - s_{\rm opt}(a_{i};\sigma|S)  \geq 
B(s_{\rm idas}(r_{i_{1}};\sigma|S),s_{\rm idas}(a_{i};\sigma|S))
-s_{\rm opt}(a_{i};\sigma|S)}\\
& = & \frac{\{b-s_{\rm idas}(r_{i_{1}};\sigma|S)\} 
s_{\rm idas}(a_{i};\sigma|S)
+\{s_{\rm idas}(a_{i};\sigma|S)-a\} 
s_{\rm idas}(r_{i_{1}};\sigma|S)}{\left\{b-s_{\rm idas}(r_{i_{1}};\sigma|S)\right\}+
\left\{s_{\rm idas}(a_{i};\sigma|S)-a\right\}}- s_{\rm opt}(a_{i};\sigma|S)\\
& = & \frac{\{b-s_{\rm idas}(r_{i_{1}};\sigma|S)\}\{s_{\rm idas}(a_{i};\sigma|S)
-s_{\rm idas}(r_{i_{1}};\sigma|S)\}}{b-a+s_{\rm idas}(a_{i};\sigma|S)-
s_{\rm idas}(r_{i_{1}};\sigma|S)}
+s_{\rm idas}(r_{i_{1}};\sigma|S)-s_{\rm opt}(a_{i};\sigma|S)\\
& \geq & \frac{\{b-s_{\rm idas}(r_{i_{1}};\sigma|S)\}d_{min}}{b-a+d_{min}}
+s_{\rm idas}(r_{i_{1}};\sigma|S)-s_{\rm opt}(a_{i};\sigma|S)\\
& = & \frac{d_{min}}{b-a+d_{min}} \left\{
b-s_{\rm idas}(r_{i_{1}};\sigma|S)+\frac{b-a+d_{min}}{d_{min}}\left(
s_{\rm idas}(r_{i_{1}};\sigma|S)-s_{\rm opt}(a_{i};\sigma|S)\right)\right\}\\
& = & \frac{d_{min}}{b-a+d_{min}}\left\{
b-s_{\rm opt}(a_{i};\sigma|S)+\frac{b-a}{d_{min}}\left(
s_{\rm idas}(r_{i_{1}};\sigma|S)-s_{\rm opt}(a_{i};\sigma|S)\right)\right\}\\
& \geq & \frac{d_{min}}{b-a+d_{min}}\left\{
b-s_{\rm opt}(a_{i};\sigma|S)+(u-1)(b-a)\right\}\\
& \geq & \frac{d_{min}}{b-a+d_{min}} 
\sum_{(a_{i},r)\in {\rm cs}_{\rm idas}^{+}(a_{i},r)}
c\left(f_{\rm bij}^{-1}(a_{i},r)|T_{\sigma}\right), 
\end{eqnarray*}
where the 1st inequality follows from (\ref{eq-2nd}). 
%
\subsection{Proof of Claim \ref{claim-bj}} \label{subsec-claim-bj}
%
For each $1 \leq j \leq \nu$, we assume that 
${\rm cs}_{\rm idas}^{+}(b_{j};\sigma) = \{(r_{j_{1}},b_{j}), \ldots,(r_{j_{v}},b_{j})\}$, 
where $j_{1},\ldots,j_{v}$ are ordered in such a way that 
\[
s_{\rm idas}(b_{j};\sigma|S) < 
s_{\rm idas}(r_{j_{1}};\sigma|S) <
s_{\rm idas}(r_{j_{2}};\sigma|S) < \cdots 
s_{\rm idas}(r_{j_{v}};\sigma|S) \leq  
s_{\rm opt}(b_{j};\sigma|S). 
\]
The rest of the proof can be shown in a way similar to the proof of 
Claim \ref{claim-ai}. 
%
\subsection{Proof of Claim \ref{claim-idas}} \label{subsec-claim-idas}
%
For the set $S$ of $k$ servers and a request sequence 
$\sigma=r_{1}\cdots r_{k}$, 
consider the case that 
$\msc{idas}_{[a,b]}$ matches $r_{j}^{h}$ with  
$s_{j}^{h}=s_{\rm idas}(r_{j}^{h};\sigma|S)$ for each $1 \leq j \leq k_{h}$. 
From the definition of $\msc{idas}_{[a,b]}$, we have that 
just before {\sc idas} matches $r_{j}^{h}$ with $s_{j}^{h}$, 
all of $s_{j}^{h}.\ldots,s_{k_{h}}^{h}$ are free and $s_{j}^{h}$ has 
the highest priority w.r.t.~$\preceq_{r_{j}^{h}}$ 
among the free servers $s_{j}^{h}.\ldots,s_{k_{h}}^{h}$. 
This is preserved to the case that 
just before $\msc{idas}_{[a,b]}$  
with the set $S_{h}$ of $k_{h}$ servers matches $r_{j}^{h}$ with 
a free server $s \in  S_{h}$ on a request sequence $\sigma_{h}$ as an input. 
Thus 
\[
s_{\rm idas}(r_{j}^{h};\sigma|S) = s_{j}^{h}=
s_{\rm idas}(r_{j}^{h};\sigma_{h}|S_{h})
\]
for each $1 \leq j \leq k_{h}$, and this completes the proof of 
the claim. 
\end{document}